%% file: byb.tex
\documentstyle[11pt]{article}
\catcode`\@=11

\@addtoreset{equation}{section}
\@addtoreset{equation}{subsection}
\def\theequation{\ifnum\value{subsection}>0\relax
\thesubsection.\arabic{equation}\relax
\else\ifnum\value{section}>0\relax
\thesection.\arabic{equation}\relax
\else\arabic{equation}\fi\fi}

\input{epic.sty}
\input{eepic.sty}

\newcommand{\cR}{{\cal R}}
\newcommand{\cS}{{\cal S}}

\begin{document}

\title{\Large \bf Boundary Yang-Baxter equation in the\\ RSOS/SOS
representation}

\author{C. Ahn\\
\small Department of Physics\\
\small Ewha Womans University\\
\small Seoul 120-750, Korea\\ \\
and \\ \\
W.M. Koo\\
\small Center for Theoretical Physics\\
\small Seoul National University\\
\small Seoul 151-742, Korea}
\date{}

\begin{titlepage}
\maketitle

\begin{abstract}
We construct and solve the boundary Yang-Baxter equation
in the RSOS/SOS representation. We find two classes of
trigonometric solutions; diagonal
and non-diagonal. As a lattice model, these two classes of solutions
correspond to RSOS/SOS models with fixed and free  boundary spins
respectively. Applied to (1+1)-dimenional quantum field theory,
these solutions give the boundary scattering amplitudes of the particles.
For the diagonal solution, we propose an algebraic Bethe ansatz method
to diagonalize the SOS-type transfer matrix with boundary and obtain the
Bethe ansatz equations.
\end{abstract}
\vspace{2cm}
\rightline{SNUTP-95-080}
\rightline{EWHA-TH-006}
\vspace{1cm}
\end{titlepage}

\section{Introduction}
In the study of the two-dimensional integrable models of quantum field
theories and statistical models, the Yang-Baxter equation (YBE)
plays essential roles in establishing the integrability and
solving the models. In the field theories, the YBE provides a consistency
condition for the two-body scattering amplitudes ($S$-matrices) in the
multi-particle scattering processes since the scattering is factorizable.
With unitarity and crossing symmetry, the YBE can determine the $S$-matrix
completely, although not uniquely due to the CDD factor which is
any function of the rapidity satisfying the unitarity and crossing symmetry
conditions.
This CDD factor is usually neglected under the the minimality assumption.
In the statistical models,
if the Boltzmann weights satisfy the YBE, row-to-row transfer matrices
with different values of the spectral parameter commute each other so that
the models are integrable.

Recently there has been a lot of efforts in extending these
approaches to models with boundaries.
The main motivation is that these models can be applied to 3D spherically
symmetric physical systems where $s$-wave element becomes dominant.
One-channel Kondo problem, monopole-catalyzed proton decay are
frequently cited examples.
Also one can generalize the conventional periodic boundary condition
of the statistical models to other types like the fixed and free conditions.

The existence of the boundary adds new quantities like
boundary scattering amplitudes and boundary Boltzmann weights,
and one needs to extend the YBE to include these objects.
The boundary Yang-Baxter equation (BYBE) (also known as the reflection
equation) \cite{cher} plays the role of the YBE for the integrable statistical
models \cite{skly,oth} and quantum field theories \cite{zam} in the
presence of a boundary; it is the necessary condition for the
integrability of these models.
The equation takes the form
\begin{equation}
R_{1}(u)S_{12}(u^{'}+u)R_2(u^{'})S_{12}(u^{'}-u)=S_{12}(u^{'}-u)S_2(u^{'})
S_{12}(u^{'}+u)R_{1}(u)
\end{equation}
where $R_{1(2)}$ is the boundary scattering (or reflection) matrix in the
auxiliary space $1(2)$ and
$S_{12}$ is the solution to the YBE.
In general, $R(u)$ need not be a ${\bf C}$-number matrix, so the
equation may be taken as the defining relation for the associative
algebra generated by the symbols $R(u)$ \cite{ryu}. This algebra
possesses a very rich
structure and has  been found to be connected with braid groups \cite{kul1},
lattice current algebra \cite{alek}, twisted Yangian \cite{naz}  and so on.
Taking the quantum space of $R(u)$ to be trivial, the BYBE is
a quadratic matrix equation which allows the matrix $R(u)$ to be solved
for given $S(u)$. To date, several solutions of the BYBE have
appeared in the form
of vertex representation, far less is known however for the
solution in the solid on solid (SOS) or restricted solid on solid (RSOS)
representation \cite{kul,pea}. Also less clear is the vertex-SOS
correspondence  associated with this algebra. In particular there is
no reason to expect the vertex-SOS transformation for the
YBE continue to hold for this algebra. Therefore finding solution in
the RSOS form may help to clarify the issue of the vertex-SOS
correspondence. Moreover, the RSOS solution will reveal the special
mathematical structure associated with this algebra when the
deformation parameter is a root of unity.

{}From a physical point of view, the solutions have applications
in statistical mechanics and field theory. In the context of statistical
mechanics, the solutions give rise to integrable SOS/RSOS models with
boundaries where the ${\bf C}$-number solution of the BYBE provides
the Boltzmann weights of the statistical models at the boundary.
The first nontrivial case gives the tri-critical Ising model. The study of
integrable statistical model with boundary will shed light on the
issue of the dependence of the Casimir energy on the boundary
and surface properties \cite{yung,bau}.
{}From the field theory point of view, the solutions are relevant to the study
of the restricted sine-Gordon model \cite{abl} and the
perturbed (coset \cite{bl,an}) conformal field theory (CFT) \cite{zam1}
with boundary. In this case, solutions to the BYBE are scattering matrices
of the particles with the boundary.

In this paper we construct the
BYBE in the RSOS/SOS representation. The
equation is studied for the diagonal and non-diagonal cases,
and the most general trigonometric solutions are found up to an
overall factor. This factor is then fixed using the boundary
crossing and unitarity conditions \cite{zam} (up to the usual CDD
ambiguity). We then construct an integrable RSOS/SOS model
using the diagonal solution and propose an algebraic Bethe ansatz method
 to diagonalize the transfer matrix of the SOS model.

\section{Solutions to the boundary Yang-Baxter equation}
\subsection{Generalities}
\begin{figure}[htbp]
\centering
\input{face1}
\caption{\label{f1} The bulk RSOS scattering matrix $S^{ab}_{dc}(u-u^{'})$}
\end{figure}
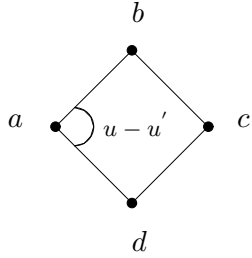

In this section we solve the BYBE for the
$\mbox{RSOS}(p)\;;$ $p=3,4\ldots$ scattering theory. The $\mbox{RSOS}(p)$
scattering theory is based on a $(p-1)$~-~fold degenerate vacuum structure,
vacua can be associated with nodes of the ${\cal A}_{p-1}$ Dynkin diagram.
The quasi particles  in the scattering theory are kinks that interpolate
neighboring vacua, they can be denoted by non-commutative symbols
$K_{ab}(u)$ where $|a-b|=1$ with $a,b=1,\ldots,p-1$ and
$u$ is related to the the kink
rapidity $\theta$ by $u=-i\theta/p$, so that the physical strip is given by
$0<{\rm Re}\;u<\pi/p$. In the rest of the paper, we will refer to $a,b$
as heights or spins.
Formally, scattering between two kinks can be represented by the following
equation (see Fig.(\ref{f1}))
\begin{equation}
K_{da}(u)K_{ab}(u^{'})=\sum_{c}S^{ab}_{dc}(u-u^{'})
K_{dc}(u^{'})K_{cb}(u)
\end{equation}
where the $S$-matrix is given by
\begin{equation}
S^{ab}_{dc}(u)={\cal U}(u)
\left(\frac{[a][c]}{[d][b]}\right)^{-u/2\gamma}W^{ab}_{dc}(u)
\label{eq:bulk}
\end{equation}
and
\begin{equation}
W^{ab}_{dc}(u)=
\left(\sin u \delta_{bd}\left(\frac{[a][c]}{[d][b]}\right)^{1/2}+
\sin(\gamma-u)\delta_{ac}\right)
\end{equation}
satisfies the YBE in the RSOS representation.

Here $[a]$ denotes the usual $q$-number given by
\[
[a]=\frac{\sin(a\gamma)}{\sin\gamma}\qquad
\gamma=\frac{\pi}{p}
\]
and the overall factor ${\cal U}(u)$ is a product of Gamma functions
satisfying the relations
\begin{eqnarray*}
{\cal U}(u){\cal U}(-u)\sin(\gamma-u) \sin(\gamma+u)&=&1\\
{\cal U}(\gamma-u)&=&{\cal U}(u)\;,
\end{eqnarray*}
and is given by
\begin{eqnarray}
{\cal U}(u)&=&\frac{1}{\pi}\Gamma\left(\frac{\gamma}{\pi}\right)
\Gamma\left(1-{u\over{\pi}}\right)
\Gamma\left(1-{\gamma\over{\pi}}+{u\over{\pi}}\right)
\prod^{\infty}_{l=1}{F_l(u)F_l(\gamma-u)\over{F_l(0)F_l(\gamma)}},\nonumber\\
F_l(u)&=&{\Gamma\left({2l\gamma\over{\pi}}-{u\over{\pi}}\right)
\Gamma\left(1+{2l\gamma\over{\pi}}-{u\over{\pi}}\right)\over{
\Gamma\left({(2l+1)\gamma\over{\pi}}-{u\over{\pi}}\right)
\Gamma\left(1+{(2l-1)\gamma\over{\pi}}-{u\over{\pi}}\right)}}\;.\label{eq:fac}
\end{eqnarray}
This factor, together with the overall $q$-number factor, ensures
that the $S$-matrix satisfies both crossing and unitarity
constraints:
\begin{eqnarray}
S^{ab}_{dc}(u)&=&S^{bc}_{ad}(\gamma-u)\\
\sum_{c^{'}}S^{ab}_{dc^{'}}(u)S^{c^{'}b}_{dc}(-u)&=&\delta_{ac}\;.
\end{eqnarray}

\begin{figure}[htbp]
\centering
\input{face2}
\caption{\label{f2} The boundary RSOS scattering matrix $R^b_{ac}(u)$}
\end{figure}
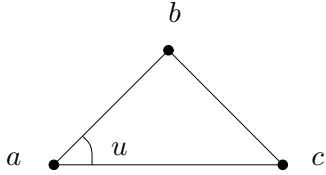

Consider now the above scattering theory in the presence of a boundary
denoted formally by ${\bf B}_{a}$, then the scattering between the kink and
the boundary is described by the equation
\begin{equation}
K_{ab}(u){\bf B}_{a}=\sum_c R^b_{ac}(u) K_{bc}(-u){\bf B}_{c}
\end{equation}
which can be given a  graphical representation shown in  Fig.(\ref{f2}). Notice
that in this representation, the boundary naturally carries an RSOS spin.

\begin{figure}[htbp]
\centering
\input{face3}
\caption{\label{f} The boundary Yang-Baxter equation}
\end{figure}
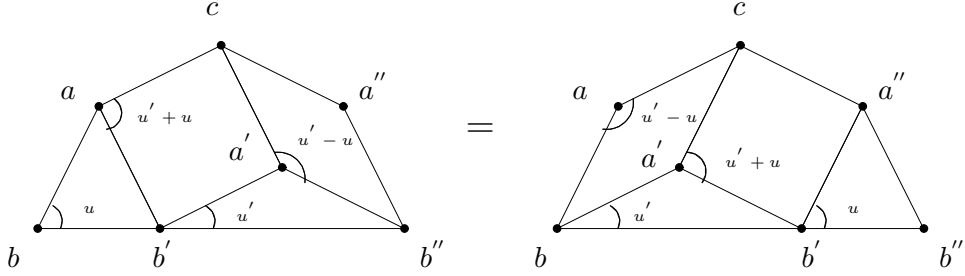

The function $R^{b}_{ac}$ is called the boundary scattering
matrix and satisfies
the BYBE, which in the RSOS representation
takes the form
\begin{eqnarray}
\lefteqn{\sum_{a^{'},b^{'}}R^a_{bb^{'}}(u)S^{ac}_{b^{'}a^{'}}(u^{'}+u)
R^{a'}_{b^{'}b^{''}}(u^{'})S^{a^{'}c}_{b^{''}a^{''}}(u^{'}-u)=}
\nonumber\\
&&
\sum_{a^{'},b^{'}}S^{ac}_{ba^{'}}(u^{'}-u)
R^{a^{'}}_{bb^{'}}(u^{'})S^{a^{'}c}_{b^{'}a^{''}}(u^{'}+u)
R^{a''}_{b^{'}b^{''}}(u)\;. \label{eq:bybe}
\end{eqnarray}
This equation is illustrated graphically in Fig.(\ref{f}).

In general, the function $R^{a}_{bc}(u)$ can be written as
\begin{equation}
R^{a}_{bc}(u)={\cal R}(u)\left(\frac{[b][c]}{[a][a]}\right)^{-u/2\gamma}
\left[\delta_{b\neq c}X^{a}_{bc}(u)+\delta_{bc}
\left\{\delta_{b,a+1}U_a(u)+\delta_{b,a-1}D_a(u)\right\}\right]\label
{eq:boundary}
\end{equation}
where ${\cal R}(u)$ has to be determined from the boundary crossing and
unitarity constraints, while $X^{a}_{bc}$ and $U_a,D_a$ have to be
determined from the BYBE.  An overall $q$-number factor has also been
multiplied to the above to cancel that from the bulk $S$-matrix
in order to simplify the BYBE.
If $X^{a}_{bc}$ does not vanish, the boundary $R$-matrix describes non-diagonal
scattering process, otherwise the scattering is called diagonal.
Note that due to the restriction
that vacuum assumes value $1,\ldots,p-1$, $X^{1}_{bc},X^{p-1}_{bc},D_1,U_{p-1}$
are not defined. The case $p=3$
has only diagonal scattering, so $X^a_{bc}$ does not exist.

\subsection{Non-diagonal scattering}
We consider the scattering where the off-diagonal component $X^a_{bc}$
is non-vanishing. To start, consider the case $b\neq c\neq b^{''}$ in
eqn.(\ref{eq:bybe}) where the BYBE gives
\begin{equation}
X^a_{a-1,a+1}(u^{'})X^{a+2}_{a+1,a+3}(u)=X^a_{a-1,a+1}(u)
X^{a+2}_{a+1,a+3}(u^{'})\;;2\leq a\leq p-4
\end{equation}
which implies that $X^{a}_{a\pm 1,a\mp 1}$ can be written as
\begin{equation}
X^{a}_{a\pm 1,a\mp 1}(u)=h_{\pm}(u)X_{\pm}^{a}
\end{equation}
where $h_{\pm}(u)$ depends only on $u$ and
$X^{a}_{\pm}$ only on $a$.

On the other hand, the case $c=b=b^{''}, a=a^{''}$ gives
\begin{equation}
X^a_{a-1,a+1}(u^{'})X^{a}_{a+1,a-1}(u)=X^a_{a-1,a+1}(u)
X^{a}_{a+1,a-1}(u^{'})\;;2\leq a\leq p-2
\end{equation}
which implies that
\begin{equation}
h_{+}(u^{'})h_{-}(u)=h_{+}(u)h_{-}(u^{'})\;,
\end{equation}
from which we conclude that
\begin{equation}
h_{+}(u)={\rm (const.)}h_{-}(u)\;.
\end{equation}
Absorbing the constant in the above equation into $X^{a}_{-}$ or $X^a_{+}$, we
can make $h_{+}$ equal to $h_{-}$ so that we can absorb
the $h_{\pm}(u)$ into the overall $\cR(u)$ factor and
treat $X^{a}_{bc}$ as $u$ independent from now on.

With this simplification, eqn.(\ref{eq:bybe}) can be broken
down into the following independent equations in addition to the above
two equations:
\begin{eqnarray}
\lefteqn{U^{'}_{a}D_{a+2}f_{+}\left(1+f_{-}{[a]\over{[a+1]}}\right)+
D^{'}_{a+2}D_{a+2}f_{-}\left(1+f_{+}{[a+2]\over{[a+1]}}\right)} \nonumber \\
&&\mbox{}+X^{a+2}_{a+1,a+3}X^{a+2}_{a+3,a+1}f_{-}=U_{a}D^{'}_{a+2}f_{+}
\left(1+f_{-}{[a+2]\over{[a+1]}}\right) \nonumber \\
&&\mbox{}+U^{'}_{a}U_{a}f_{-}\left(1+f_{+}{[a]\over{[a+1]}}\right)
+X^{a}_{a-1,a+1}X^{a}_{a+1,a-1}f_{-}\label{eq:s0}
\end{eqnarray}
for $1\leq a\leq p-3$,
\begin{eqnarray}
&&D^{'}_{a+1}f_{-}\left(1+f_{+}{[a+1]\over{[a]}}\right)+
U^{'}_{a-1}f_{+}\left(1+f_{-}{[a-1]\over{[a]}}\right)\nonumber\\
&&\mbox{\hspace{1cm}}=U_{a-1}f_{+}-U_{a+1}f_{-}\label{eq:s1a}\\
&&\nonumber\\
&&U^{'}_{a}f_{-}\left(1+f_{+}{[a]\over{[a+1]}}\right)+
D^{'}_{a+2}f_{+}\left(1+f_{-}{[a+2]\over{[a+1]}}\right)\nonumber\\
&&\mbox{\hspace{1cm}}=D_{a+2}f_{+}-D_{a}f_{-}\label{eq:s1b}
\end{eqnarray}
for $2\leq a\leq p-3$, and
\begin{eqnarray}
&&U^{'}_{a-2}f_{+}f_{-}{[a][a-2]\over{[a-1]^2}}-U^{'}_{a}
+D^{'}_{a}\left(1+f_{-}{[a]\over{[a-1]}}\right)
\left(1+f_{+}{[a]\over{[a-1]}}\right)\nonumber\\
&&\mbox{\hspace{1cm}}=D_{a}\left(1+f_{+}{[a]\over{[a-1]}}\right)
-U_{a}\left(1+f_{-}{[a]\over{[a-1]}}\right)
\label{eq:s2a}\\&&\nonumber \\
&&D^{'}_{a+2}f_{+}f_{-}{[a][a+2]\over{[a+1]^2}}-D^{'}_{a}+
U^{'}_{a}\left(1+f_{-}{[a]\over{[a+1]}}\right)
\left(1+f_{+}{[a]\over{[a+1]}}\right)\nonumber\\
&&\mbox{\hspace{1cm}}=U_{a}\left(1+f_{+}{[a]\over{[a+1]}}\right)
-D_{a}\left(1+f_{-}{[a]\over{[a+1]}}\right)\label{eq:s2b}
\end{eqnarray}
for $2\leq a\leq p-2$. The last four equations are derived based on
the assumption that the off-diagonal weight $X^a_{bc}$ is {\bf nonvanishing}.
In the above equations, we used a compact notation
where
$U_a=U_a(u),\ U^{'}_a=U_a(u^{'})$ (similarly for $D_a$)
and
\[f_{\pm}=\sin(u^{'}\pm u)/\sin(\gamma-u^{'}\mp u)\;.\]

In addition, it should also be mentioned that the last term
in the rhs (lhs) of eqn.(\ref{eq:s0}) is present only when $a\neq 1 (p-3)$
and the first terms of eqns.(\ref{eq:s2a}) and (\ref{eq:s2b}) are
allowed only for $a\neq 2$ and $a\neq p-2$, respectively. Let us call these
terms that are not supposed to be there the ``unwanted'' terms.
The way to solve these equations is to construct some recursion
relations for the functions $U_a,D_a,X^a_{a\pm 1,a\mp 1}$ and
solve the recursion relations subject to the conditions
that the above mentioned ``unwanted'' terms are zero.
Explicitly, one has
\begin{eqnarray}
X^{1}_{0,2}X^{1}_{2,0}&=X^{p-1}_{p-2,p}X^{p-1}_{p,p-2}&=0\;,\label{eq:bc}\\
U_0&=D_{p}&=0\;.\label{eq:bc2}
\end{eqnarray}
A few comments are in order here. Notice that the coefficients of $U_0$ and
$D_p$ in eqns.(\ref{eq:s2a}), (\ref{eq:s2b}) are $[0]$ and $[p]$ respectively,
which vanish by construction. So in principle, one needs not impose the above
condition, eqn.(\ref{eq:bc2}), for the recursion relations of
$U_a,D_a$. However, we shall see later that a particular solution of
$U_a,D_a$ is given by
\begin{equation}
U_a=-D_a={\rm (const.)}\frac{1}{[a]}
\end{equation}
which cancels the vanishing coefficients at $a=0, p$, and renders the
unwanted terms nonvanishing. Therefore, we have to impose eqn.(\ref{eq:bc2})
on the above solution.

Notice also that most of the above equations do not apply to the case
$p=4$, so we shall deal with this case separately.

{}From the above it is clear that eqns.(\ref{eq:s0})-(\ref{eq:bc}) can be
divided into two parts;
eqn.(\ref{eq:s1b})-eqn.(\ref{eq:s2b}) determine $U_a$ and $D_a$ while
eqns.(\ref{eq:s0}),(\ref{eq:bc}) determine $X^{a}_{bc}$.
Indeed, comparing eqn.(\ref{eq:s1a}) with
eqn.(\ref{eq:s2a}) and similarly eqn.(\ref{eq:s1b}) with
eqn.(\ref{eq:s2b}), we deduce that
\begin{eqnarray}
(D_a^{'}-D_a)\left(1+f_{+}\frac{[a]}{[a-1]}\right)&=&(U^{'}_{a-2}-U_{a-2})
f_{+}\frac{[a]}{[a-1]}\nonumber\\
&&\mbox{}+(U_a^{'}-U_a)\\
(U_a^{'}-U_a)\left(1+f_{+}\frac{[a]}{[a+1]}\right)&=&(D^{'}_{a+2}-D_{a+2})
f_{+}\frac{[a]}{[a+1]}\nonumber\\
&&\mbox{}+(D_a^{'}-D_a)\;.
\end{eqnarray}
Substituting one into another, we get
\begin{eqnarray}
\frac{(U^{'}_{a+2}-U_{a+2})-(U^{'}_{a}-U_{a})}{(U^{'}_{a}-U_{a})
-(U^{'}_{a-2}-U_{a-2})}&=&\frac{\sin\left((a+1)\gamma+u^{'}+u\right)}
{\sin\left((a-1)\gamma+u^{'}+u\right)}\label{eq:rec1}\\
\frac{(D^{'}_{a+2}-D_{a+2})-(D^{'}_{a}-D_{a})}{(D^{'}_{a}-D_{a})
-(D^{'}_{a-2}-D_{a-2})}&=&\frac{\sin\left((a+1)\gamma-u^{'}-u\right)}
{\sin\left((a-1)\gamma-u^{'}-u\right)}\label{eq:rec2}
\end{eqnarray}
and writing the rhs respectively as
\begin{eqnarray*}
&\frac{\textstyle\cos\left(2u^{'}+(a+1)\gamma\right)
-\cos\left(2u+(a+1)\gamma\right)}
{\textstyle\cos\left(2u^{'}+(a-1)\gamma\right)
-\cos\left(2u+(a-1)\gamma\right)}&\\
&\frac{\textstyle\cos\left(2u^{'}-(a+1)\gamma\right)
-\cos\left(2u-(a+1)\gamma\right)}
{\textstyle\cos\left(2u^{'}-(a-1)\gamma\right)
-\cos\left(2u-(a-1)\gamma\right)}&\;,
\end{eqnarray*}
it is clear that
\begin{eqnarray*}
U_{a+2}(u)-U_{a}(u)&=&\cos\left(2u+(a+1)\gamma\right)+\beta^{'}_a\\
D_{a+2}(u)-D_{a}(u)&=&-\cos\left(2u-(a+1)\gamma\right)+\phi^{'}_a
\end{eqnarray*}
where $\beta^{'}_a,\phi^{'}_a$ are unknown functions of $a$ only.
Iterating the above, one finds
\begin{eqnarray}
U_a(u)&\propto&\sin(2u+a\gamma)+\alpha(u)+\beta_a\\
D_a(u)&\propto&\sin(2u-a\gamma)+\gamma(u)+\phi_a
\end{eqnarray}
where $\alpha(u),\gamma(u)$ are unknown functions of $u$,
and $\beta_a,\phi_a$ of $a$.
Furthermore, from eqns.(\ref{eq:s0})-(\ref{eq:s2b}), one can establish
the following symmetry
\begin{equation}
U_a(u)=-D_a(-u)\;,\label{eq:sym}
\end{equation}
which reduces the number of unknown functions to two, namely,
$\alpha(u)$ and $\beta_a$.
To determine them, we have to substitute the above expressions
for $U_a,D_a$ back into eqns.(\ref{eq:s1a})-(\ref{eq:s2b}).
Notice, however, that
these equations are linear in $U_a,D_a$, so it suffices to consider
$\alpha(u)$ and $\beta_a$ separately. Doing this amounts to finding
special solutions to eqns.(\ref{eq:s1a})-(\ref{eq:s2b}) where
$U_a,D_a$ have only $u$ or $a$ dependence. The solutions are given by
\begin{equation}
\alpha(u)={\rm(const.)}\frac{1}{\sin(2u)}\;,\hspace{2cm}\beta_a={\rm(const.)}\frac{1}{\sin(a\gamma)}
\end{equation}
respectively.
Imposing eqn.(\ref{eq:bc2}) on $\alpha(u),\beta_a$, one finds
\begin{equation}
\beta_a=0\;.
\end{equation}
So far the above $U_a,D_a$ are obtained based on the assumption that
the numerators and denominators of the lhs of
eqns.(\ref{eq:rec1}),(\ref{eq:rec2}) are nonvanishing. In fact, there are
additional solution to these recursion relations where the above mentioned
numerators and denominators vanish;
\begin{eqnarray*}
U_{a+2}(u^{'})-U_{a}(u^{'})-U_{a+2}(u)+U_{a}(u)&=&0\\
D_{a+2}(u^{'})-D_{a}(u^{'})-D_{a+2}(u)+D_{a}(u)&=&0\;.
\end{eqnarray*}
The additional solution is given by
\begin{equation}
U_a(u)={\rm(const.)}\frac{\sin(a\gamma+2u)}{\sin(2u)\sin(a\gamma)}\;,
\hspace{1cm}
D_a(u)={\rm(const.)}\frac{\sin(a\gamma-2u)}{\sin(2u)\sin(a\gamma)}\;.
\end{equation}
However, on imposing the condition eqn.(\ref{eq:bc2}), one finds
that they have to be zero.
In summary, the general non-diagonal solution to the four linear
equations is given by
\begin{equation}
\begin{array}{lcl}
U_a(u)&=&\sin(2u+a\gamma)+{\rm (const.)}\frac{\textstyle 1}
{\textstyle \sin(2u)}
\\
D_{a+1}(u)&=&\sin\left(2u-(a+1)\gamma\right)+{\rm (const.)}\frac{\textstyle 1}
{\textstyle \sin(2u)}
\end{array}\label{eq:udn}
\end{equation}
where $1\leq a\leq p-2$.

Having found $U_a,D_a$, the function $X^a_{bc}$ can be
easily obtained from eqn.(\ref{eq:s0}),
which can be further simplified with the symmetry properties given
in eqn.(\ref{eq:sym}) and taking $u^{'}$ to be $-u$ since $X^a_{bc}$
does not depend on the rapidity. This gives
\[X^{a+2}_{a+1,a+3}X^{a+2}_{a+3,a+1}-X^{a}_{a-1,a+1}X^{a}_{a+1,a-1}
=D_{a+2}(u)U_{a+2}(u)-D_{a}(u)U_a(u)\;.
\]
Substituting $U_a,D_a$ into the rhs and iterating the equations, we get
\begin{equation}
X^{a}_{a-1,a+1}X^{a}_{a+1,a-1}=\sin^2\gamma-\sin^2(a\gamma)\label{eq:xn}
\end{equation}
where the use of eqn.(\ref{eq:bc}) forces the first term to be
$\sin^2\gamma$ and the coefficient of $\frac{1}{\sin(2u)}$ to vanish.
Since this equation determines only the product, $X^{a}_{a-1,a+1}$
and $X^{a}_{a+1,a-1}$ are determined up to a gauge factor.

These solutions have the property that
\begin{equation}
U_{p-a}(u)=-D_a(u)\;,\hspace{1cm}X^{p-a}_{p-a-1,p-a+1}
X^{p-a}_{p-a+1,p-a-1}=X^{a}_{a-1,a+1}X^{a}_{a+1,a-1}\;.
\end{equation}

Next,  we consider the BYBE for $p=4$.
The functions $U_2,D_2$ and $X^2_{13},X^2_{31}$ satisfy the following equations
\begin{eqnarray}
X^{2}_{13}(u^{'})X^2_{31}(u)&=&X^{2}_{13}(u)X^{2}_{31}(u^{'})\label{eq:ss0}\\
\lefteqn{U_{2}\left(1+\sqrt{2}f_{-}\right)+D_{2}^{'}\left(1+\sqrt{2}
f_{+}\right)\left(1+\sqrt{2}f_{-}\right)}\hspace{3cm}\nonumber\\
&=&U_2^{'}+D_{2}\left(1+\sqrt{2}f_{+}\right)\\
\lefteqn{D_{2}\left(1+\sqrt{2}f_{-}\right)+U_{2}^{'}\left(1+\sqrt{2}
f_{+}\right)\left(1+\sqrt{2}f_{-}\right)}\hspace{3cm}\nonumber\\
&=&D_2^{'}+U_{2}\left(1+\sqrt{2}f_{+}\right)\;.
\end{eqnarray}
The functions $U_1, D_3$ are diagonal scattering components and
do not couple to the above equations and we shall defer to next
section for their computation.
Here we have used the compact notation introduced earlier for
$U_a, D_a$, and written out explicitly the rapidity dependence of $X^a_{bc}$.
As before, the last two equations have been derived based on the assumption
that $X^2_{13},X^2_{31}$ are nonvanishing. From eqn.(\ref{eq:ss0}), we deduce
that
\begin{equation}
X^2_{13}(u)={\rm (const.)}X^2_{31}(u)\;.
\end{equation}
We are free to take $X^2_{31}$ as unity and the above implies that
$X^2_{13}$ is just a gauge factor, which we call $g$.

The rest of the equations can be turned into ordinary
differential equations in the limit $u^{'}\rightarrow u$,
giving
\begin{eqnarray}
\left(\dot{U}_2(u)+\dot{D}_2(u)\right)\tan(2u)+2\left(U_2(u)
+D_2(u)\right)&=&0\\
\left(\dot{U}_2(u)-\dot{D}_2(u)\right)\cot(2u)-2\left(U_2(u)
-D_2(u)\right)&=&0\;,
\end{eqnarray}
which can be integrated to give
\begin{eqnarray}
U_2(u)&=&B/\sin(2u)+C\cos(2u)\\
D_2(u)&=&B/\sin(2u)-C\cos(2u)\;,
\end{eqnarray}
with $B,C$ as free parameters.

This completes the determination of the non-diagonal solutions of the BYBE.

\subsection{Diagonal scattering}

For the diagonal scattering, we take
\begin{equation}
R^{a}_{bc}(u)=([b]/[a])^{-u/\gamma}{\cal R}(u)
\delta_{bc}\left[\delta_{b,a+1}U_a(u)+\delta_{b,a-1}D_a(u)\right]\;,
\end{equation}
and the BYBE is equivalent to a single equation
\begin{eqnarray}
\lefteqn{U_{a-1}(u^{'})D_{a+1}(u)\sin(u^{'}+u)
\sin(a\gamma-u^{'}+u)}\nonumber\\
&&\mbox{}+D_{a+1}(u^{'})D_{a+1}(u)\sin(u^{'}-u)
\sin(a\gamma+u^{'}+u)=\nonumber\\
\lefteqn{U_{a-1}(u)D_{a+1}(u^{'})\sin(u^{'}+u)
\sin(a\gamma+u^{'}-u)}\nonumber\\
&&\mbox{}+U_{a-1}(u^{'})U_{a-1}(u)\sin(u^{'}-u)
\sin(a\gamma-u^{'}-u)\;,\label{eq:dia}
\end{eqnarray}
which holds only for $2\leq a\leq p-2$. So the functions $U_{p-2}$
and $D_2$ can not be determined from the BYBE.
The above can be recast into
\begin{eqnarray}
\lefteqn{[\cos(a\gamma+2u)Z(u)-\cos(a\gamma-2u)][Z(u^{'})-1]=}\nonumber\\
&&\mbox{ }[\cos(a\gamma+2u^{'})Z(u^{'})-\cos(a\gamma-2u^{'})][Z(u)-1]
\end{eqnarray}
where
\[Z_a(u)\equiv D_{a+1}(u)/U_{a-1}(u)\;.\]
One can easily find that the general solution is
\begin{equation}
\frac{D_{a+1}(u)}{U_{a-1}(u)}=\frac{\sin(\xi_a+u)\sin(\xi_a+a\gamma-u)}
{\sin(\xi_a-u)\sin(\xi_a+a\gamma+u)}\label{eq:ratio}
\end{equation}
where $\xi_a$ is a free parameter.

Thus for the diagonal solution, there are $p-3$ parameters $\xi_a$.
This solution gives $p-1$ distinct diagonal scattering theories of
kinks, each with a specific boundary ${\bf B}_a$. There is one free
parameter $\xi_a$ for each theory, except for the cases of
$a=1,p-1$ where there is no free parameter.

This solution includes a particular case of $p=4$ which has
been omitted earlier.

Further relations from boundary unitarity and crossing symmetry will be
required to disentangle $U_{a-1}$ and $D_{a+1}$, and determine $U_2$
and $D_{p-2}$, see later.

\subsection{Boundary unitarity and crossing symmetry}

The boundary unitarity and crossing symmetry conditions of the scattering
matrix $R^a_{bc}(u)$ determine to some extent the overall factor $R(u)$.
These conditions can be written as
\begin{eqnarray}
\sum_{c}R^{a}_{bc}(u)R^{a}_{cd}(-u)&=&\delta_{bd}\\
\sum_{d}S^{ac}_{bd}(2u)R^{d}_{bc}(\gamma/2+u)&=&R^{a}_{bc}(\gamma/2-u)\;.
\end{eqnarray}

Consider the non-diagonal scattering ($p>4$) first.
Substituting the expression for  $R^a_{bc}$ into the unitarity
condition, we get the following
\[{\cal R}(u){\cal R}(-u)\left[X^a_{bd}X^a_{db}\delta_{b\neq d}
-U_a(u)D_a(u)\right]=1\]
where use has been made of the symmetry eqn.(\ref{eq:sym}). Applying the
results eqns.(\ref{eq:udn}), (\ref{eq:xn}) to the above leads to
\begin{equation}
{\cal R}(u){\cal R}(-u)\left(\sin^2\gamma-4\sin^2u+4\sin^4u\right)=1\;.
\label{eq:r1}
\end{equation}
While for crossing symmetry condition we get
\begin{equation}
{\cal U}(2u){\cal R}(\gamma/2+u)\sin(\gamma-2u)=
{\cal R}(\gamma/2-u)\label{eq:r2}
\end{equation}
using the relations
\begin{eqnarray}
&&D_{a+2}(\gamma-u){[a+2]\over{[a+1]}}+
U_a(\gamma-u){[a]\over{[a+1]}}\nonumber\\
&&\hspace{2cm}=f(2u)\left(U_a(u)-U_a(\gamma-u)\right)\\
&&U_a(\gamma-u){[a]\over{[a+1]}}+D_{a+2}(\gamma-u)
{[a+2]\over{[a+1]}}\nonumber\\
&&\hspace{2cm}=f(2u)\left(D_{a+2}(u)-D_{a+2}(\gamma-u)\right)
\end{eqnarray}
which are obtained from eqn.(\ref{eq:s1b}) in the limit
$u^{'}\rightarrow \gamma-u$. Here $f(u)=\sin u /\sin(\gamma-u)$.

The factor ${\cal R}(u)$ can be determined from eqns.(\ref{eq:r1}),
(\ref{eq:r2}) up to the usual CDD ambiguity by separating
${\cal R}(u)={\cal R}_0(u){\cal R}_1(u)$ where ${\cal R}_0$ satisfies
\begin{equation}
\begin{array}{lll}
\cR_{0}(u)\cR_0(-u)&=&1\\
{\cal U}(2u)\cR_0(\gamma/2+u)\sin(\gamma-2u)&=&\cR_0(\gamma/2-u)\;,
\end{array}
\end{equation}
whose minimal solution reads
\begin{equation}
R_0(u)=\frac{F_0(u)}{F_0(-u)}
\end{equation}
where $F_l(u)$ has been given in eqn.(\ref{eq:fac}). While $\cR_1$ satisfies
\begin{equation}
\begin{array}{ll}
&\cR_1(u)\cR_1(-u)\left(\sin^2\gamma-4\sin^2u+4\sin^4u\right)=1\\
&\cR_1(u)=\cR_1(\gamma-u)
\end{array}
\end{equation}
with minimum solution
\begin{equation}
\cR_1(u)=\frac{1}{2}\sigma(\frac{\gamma}{2},u)\sigma(\frac{\pi-\gamma}{2},u)
\end{equation}
where
\begin{equation}
\begin{array}{lcl}
\sigma(x,u)&=&\frac{\textstyle\prod(x,\frac{\gamma}{2}-u)
\prod(-x,\frac{\gamma}{2}-u)
\prod(x,-\frac{\gamma}{2}+u)\prod(-x,-\frac{\gamma}{2}+u)}
{\textstyle\prod^2(x,\frac{\gamma}{2})\prod^2(-x,\frac{\gamma}{2})}\\&&\\
\prod(x,u)&=&\displaystyle\prod_{l=0}^{\infty}
\frac{\textstyle\Gamma\left(\frac{1}{2}+(2l+\frac{1}{2})\frac{\gamma}{\pi}
+\frac{x}{\pi}-\frac{u}{\pi}\right)}
{\textstyle\Gamma\left(\frac{1}{2}+(2l+\frac{3}{2})
\frac{\gamma}{\pi}+\frac{x}{\pi}-\frac{u}{\pi}\right)}\;.
\end{array}
\end{equation}
For $p=4$, the crossing symmetry condition is the same as eqn.(\ref{eq:r2}),
but unitarity now requires that
\begin{equation}
\cR(u)\cR(-u)\left(g+C^2\cos^2(2u)-B^2/\sin^2(2u)\right)=1
\end{equation}
These equations can again be solved by separation (see \cite{meahn}
for details).
It should be remarked that there are two parameters in this case while
there is only one in the higher $p$ cases.

Next, we consider the diagonal case. For convenience, we
set the undefined terms $U_{0},D_{p}$ to be zero. Unitarity and
crossing symmetries relations give respectively
\begin{equation}
\begin{array}{lll}
\cR(u)\cR(-u)U_a(u) U_a(-u)&=&1\\
\cR(u)\cR(-u)D_{a+1}(u) D_{a+1}(-u)&=&1
\end{array}
\end{equation}
and
\begin{equation}
\begin{array}{c}
{\cal U}(\gamma-2u)\cR(\gamma-u)
\left[U_a(\gamma-u)\sin\gamma\sin(a\gamma+2u)\right.\\
\mbox{ }\left.+D_{a+2}(\gamma-u)\sin(\gamma-2u)
\sin\left((a+2)\gamma\right)\right]
=\cR(u)U_a(u)\sin((a+1)\gamma)\\ \\
{\cal U}(\gamma-2u)\cR(\gamma-u)
\left[D_{a+1}(\gamma-u)\sin\gamma\sin\left((a+1)\gamma-2u\right)\right.\\
\mbox{ }\left.+U_{a-1}(\gamma-u)\sin(\gamma-2u)
\sin\left((a-1)\gamma\right)\right]
=\cR(u)D_{a+1}(u)\sin(a\gamma)
\end{array}
\end{equation}
for $ 1\leq a\leq p-2$.


These equations can be solved separately; we set
\begin{equation}
\begin{array}{lll}
\cR(u)\cR(-u)&=&1\\
{\cal U}(2u)\cR(\gamma/2+u)\sin(\gamma-2u)&=&\cR(\gamma/2-u)\;,
\end{array}
\end{equation}
so that $\cR(u)$ has exactly the same solution as that of $\cR_0(u)$
considered earlier. While $U_a,D_a$ satisfy
\begin{equation}
\begin{array}{lll}
U_a(u)U_a(-u)&=&1\\
D_{a+1}(u) D_{a+1}(-u)&=&1
\end{array}\;\label{eq:u1}
\end{equation}
and
\begin{equation}
\begin{array}{l}
U_a(\gamma-u)\sin\gamma\sin(a\gamma+2u)+D_{a+2}(\gamma-u)
\sin(\gamma-2u)\sin((a+2)\gamma)\\
\mbox{\hspace{2cm}}=U_a(u)\sin(2u)\sin((a+1)\gamma)\;,\\
D_{a+1}(\gamma-u)\sin\gamma\sin((a+1)\gamma-2u)+U_{a-1}(\gamma-u)
\sin(\gamma-2u)\sin((a-1)\gamma)\\
\mbox{\hspace{2cm}}=D_{a+1}(u)\sin(2u)\sin(a\gamma)\;.
\end{array}
\end{equation}
for $1\leq a\leq p-2$.
Substituting eqn.(\ref{eq:ratio}) into the above
we get a relation between $U_a(u)$
($D_a(u)$) and $U_a(\gamma-u)$ ($D_a(\gamma-u)$);
\begin{equation}
\begin{array}{lll}
\frac{\textstyle U_{a-1}(u)}{\textstyle U_{a-1}(\gamma-u)}&=&
\frac{\textstyle \sin(2(\gamma-u))\sin(\xi_a-u)\sin(\xi_a+a\gamma+u)
}{\textstyle \sin(2u)\sin(\xi_a-(\gamma-u))\sin(\xi_a+a\gamma+(\gamma-u))}\\
&&\\
\frac{\textstyle D_{a+1}(u)}{\textstyle D_{a+1}(\gamma-u)}&=&
\frac{\textstyle \sin(2(\gamma-u))\sin(\xi_a+u)
\sin(\xi_a+a\gamma-u)}{\textstyle \sin(2u)
\sin(\xi_a+(\gamma-u))\sin(\xi_a+a\gamma-(\gamma-u))}\;.
\end{array}\label{eq:diag2}
\end{equation}
These relations together with eqn.(\ref{eq:u1}) give the minimal
solutions of $U_a(u)$ and $D_a(u)$
\begin{eqnarray}
U_{a-1}(u)&=&\frac{1}{2}\sin 2(\gamma-u)\sin(\xi_a-u)\sin(\xi_a+a\gamma+u)
\sigma(\gamma,u)\nonumber\\
&&\mbox{}\times\sigma(\frac{\pi}{2}-\gamma,u)
\sigma(\frac{\pi}{2}-\xi_a,u)\sigma(\frac{\pi}{2}-\xi_a-a\gamma,u)
\label{eq:udsol1}\;,\\
D_{a+1}(u)&=&\frac{1}{2}\sin 2(\gamma-u)\sin(\xi_a+u)\sin(\xi_a+a\gamma-u)
\sigma(\gamma,u)\nonumber\\
&&\mbox{}\times\sigma(\frac{\pi}{2}-\gamma,u)
\sigma(\frac{\pi}{2}-\xi_a,u)\sigma(\frac{\pi}{2}-\xi_a-a\gamma,u)\;.
\label{eq:udsol2}
\end{eqnarray}
for $2\leq a\leq p-2$ and
\begin{eqnarray}
U_{p-2}(u)&=&\frac{1}{2}\sin 2(\gamma-u)\sigma(\gamma,u)
\sigma(\frac{\pi}{2}-\gamma,u)\;,\\
D_{2}(u)&=&\frac{1}{2}\sin 2(\gamma-u)\sigma(\gamma,u)
\sigma(\frac{\pi}{2}-\gamma,u)\;.
\end{eqnarray}

To summarize, there are two classes of solutions to the BYBE;
diagonal and non-diagonal. Unlike the solution in vertex representation,
the former is not a special limit of the latter.
In fact, the diagonal solution carries $p-3$ free parameters
while nondiagonal has non.

\subsection{Comments on the SOS model}

We have considered from the start heights take values  from $1$ to
$p-1$, which is necessary for the bulk scattering weights to be finite as
the parameter $\pi/\gamma=p$ is a positive  integer. When $\pi/\gamma$
is not a rational number, there is no bounds on the heights and the
corresponding representation is known as solid-on-solid (SOS).
The removal of the heights'
restriction certainly affects the solution of the BYBE.

For the diagonal solution,
it is clear that essentially there is no difference
between the SOS and RSOS solution, where the SOS solution is given
by eqns.(\ref{eq:udsol1}),(\ref{eq:udsol2}) for any integer
$a$ and the overall factor $\cR(u)$ is the same as before.
Hence there is a free parameter
for each height.

While for the non-diagonal solution,
the conditions eqns.(\ref{eq:bc}),(\ref{eq:bc2})
of the recursion relations do not apply at
all, thus $\beta_a\neq 0$ and the corresponding diagonal weights are given by
\begin{equation}
\begin{array}{lcl}
U_a(u)&=&c_1\sin(2u+a\gamma)+c_2\frac{\textstyle 1}{\textstyle \sin(a\gamma)}
+c_3\frac{\textstyle \sin(a\gamma+2u)}{\textstyle \sin(2u)\sin(a\gamma)}
+c_4\frac{\textstyle 1}{\textstyle \sin(2u)}\\ \\
D_a(u)&=&c_1\sin(2u-a\gamma)-c_2\frac{\textstyle 1}{\textstyle \sin(a\gamma)}
+c_3\frac{\textstyle \sin(a\gamma-2u)}{\textstyle \sin(2u)\sin(a\gamma)}
+c_4\frac{\textstyle 1}{\textstyle \sin(2u)}
\end{array}
\end{equation}
where $c_i\;;i=1,\ldots,4$ are free parameters. The off-diagonal
weights satisfy
\begin{eqnarray}
X^{a}_{a-1,a+1}X^{a}_{a+1,a-1}&=&c^2_0+2c_1c_4\cos(a\gamma)
-c_1^2\sin^2(a\gamma)-c_2^2\frac{1}{\sin^2(a\gamma)}   \nonumber\\
&&\mbox{}-2c_2c_3\frac{\cos(a\gamma)}{\sin^2(a\gamma)}
-c_3^2\frac{1}{\sin^2(a\gamma)}
\label{eq:xn1}\;
\end{eqnarray}
and are determined up an additive constant $c_0$.
Of these five free parameters one is an overall factor, so there are
four free parameters in this case. The overall
factor $\cR_0$ is given as before, but $\cR_1(u)$ now contains all the
information of the boundary conditions and has to satisfy
\begin{equation}
\begin{array}{l}
\cR_1(u)\cR_1(-u)\left(c_0^2-c_1^2\sin^2(2u)+2c_1c_2\cos(2u)-
\frac{\textstyle c_3^2+2c_3c_4\cos(2u)+c_4^2}{\textstyle\sin^2(2u)}\right)=1\\
\cR_1(u)=\cR_1(\gamma-u)
\end{array}
\end{equation}
whose minimum solution is
\begin{equation}
\cR_1(u)=\frac{2\sigma(\vartheta_1,u)\sigma(\vartheta_2,u)
\sigma(\vartheta_3,u)\sigma(\vartheta_4,u)
}{(c_3+c_4)\sigma(0,u)}\cS(u)
\end{equation}
where $\vartheta_i$ are related to $c_i$ via
\begin{eqnarray*}
\prod_{i=1}^4\cos^2\vartheta_1&=&\left(\frac{c_3+c_4}{4c_1}\right)^2\;,\\
\sum_{i<j<k=1}^4\cos^2\vartheta_i\cos^2\vartheta_j\cos^2\vartheta_k
&=&\frac{c_0^2+c_3c_4+2c_1c_2}{4c_1^2}\;,\\
\sum_{i<j=1}^4\cos^2\vartheta_i\cos^2\vartheta_j
&=&\frac{c_0^2+5c_1c_2+4c_1^2}{4c_1^2}\;,\\
\sum_{i=1}^4\cos^2\vartheta_i&=&\frac{c_1c_2+2c_1^2}{c_1^2}
\end{eqnarray*}
and $\sigma(x,u)$ has been given before, while
\begin{equation}
\cS(u)=\prod_{l=0}^{\infty}\frac{\Gamma\left((2l+1)\frac{\gamma}{\pi}
+\frac{u}{\pi}\right)\Gamma\left((2l+2)\frac{\gamma}{\pi}+\frac{u}{\pi}\right)
\Gamma\left(1-(2l+1)\frac{\gamma}{\pi}-\frac{u}{\pi}\right)
\Gamma\left(1-(2l+2)\frac{\gamma}{\pi}-\frac{u}{\pi}\right)}
{\Gamma\left((2l+1)\frac{\gamma}{\pi}-\frac{u}{\pi}\right)
\Gamma\left((2l+2)\frac{\gamma}{\pi}-\frac{u}{\pi}\right)
\Gamma\left(1-(2l+1)\frac{\gamma}{\pi}+\frac{u}{\pi}\right)
\Gamma\left(1-(2l+2)\frac{\gamma}{\pi}+\frac{u}{\pi}\right)}\;.
\end{equation}
It is interesting to point out that the $c_2,c_3,c_4$ terms in $U_a,D_a$
are related to the boundary $K$-matrix of the 6-vertex model\cite{gon} via
the vertex-SOS intertwiner\cite{bax,zou,men}. In this light, one can
regard the term $c_1$ to be the solution special to the SOS/RSOS
representation of the BYBE. In fact, it is not clear whether there
exist an intertwiner that relate it to the $K$-matrix of the 6-vertex
model. It is then natural to wonder if similar special solution
occur in the elliptic case\cite{hou}.

Also intriguing is that restricting SOS to RSOS corresponds to
\begin{equation}
c_2=c_3=c_4=0,\qquad c_0=c_1\sin(a\gamma)\;.
\end{equation}

\section{Commuting transfer matrix}

Following the technique proposed in \cite{skly} for the vertex model,
one can similarly construct
a family of commuting transfer matrix for the RSOS/SOS model with boundary.

To start, it can be shown that if $R^{a}_{bc}$ is a solution to the
BYBE in the RSOS/SOS form then the following combination
\begin{equation}
\sum_a S^{fe}_{ba}(u-u_1)S^{ae}_{cd}(u+u_1)R^a_{bc}(u)
\end{equation}
also  satisfies the BYBE, where $S^{fe}_{ba}(u)$ is the solution of the
bulk YBE given in eqn.(\ref{eq:bulk}) and $u_1$ is an arbitrary parameter.
The proof is essentially the same as the vertex case
given in \cite{skly} and we shall not repeat it here.

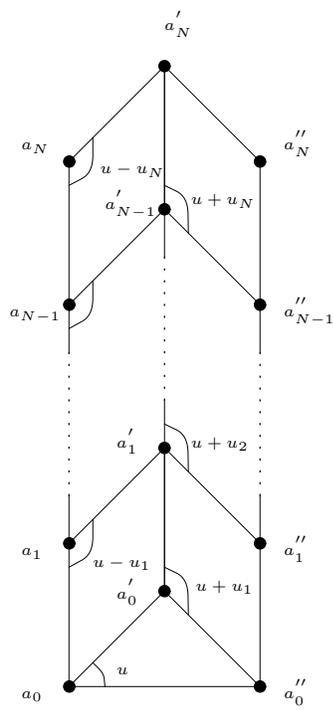
\begin{figure}[htbp]
\centering
\input{face4}
\caption{\label{f4} ``Decorated" boundary scattering matrix}
\end{figure}

It is convenient
to think of the BYBE as the defining relation of some associative
algebra generated by the symbol $R^a_{bc}$. So the solutions given in  the
previous section correspond to particular representations of this
algebra where the ``quantum space'' is trivial and the auxiliary space
is the space of a one-step path ${\cal P}_1$ on a truncated Bratteli
diagram with $ab$ and $ac$ being in- and out-state, respectively.
In this context, the above ``decorated'' solution then corresponds
to a representation
whose quantum space is isomorphic to ${\cal P}_1$ that is formed by
the nodes $f,b$ ($d,c$). Clearly, the above construction can be
repeated for an arbitrary number
of times (say $N+1$) giving a boundary $R$-matrix that acts on ${\cal P}_{N}$,
the collection of $N$-step paths on a truncated Bratteli diagram.
We shall denote such a solution as ${\bf R}^{a}_{bc}$ which should be
regarded as an operator acting on ${\cal P}_{N}$.
Its matrix element is explicitly given by
\begin{eqnarray}
{\bf R}^a_{bc}(u)_{a_0,\cdots,a_{N};a^{''}_0,\cdots,
a^{''}_{N}}&=&\delta_{aa^{'}_N}\delta_{b a_N}
\delta_{ca^{''}_N}\prod_{i=1}^N\sum_{a^{'}_{i-1}}
\left( S^{a_ia^{'}_i}
_{a_{i-1}a^{'}_{i-1}}(u-u_i)\right.\nonumber\\
&&{}\times\left.S^{a^{'}_{i-1}a^{'}_i}_{a^{''}_{i-1}
a^{''}_{i}}(u+u_i)\right)R^{a^{'}_{0}}_{a_{0}
a^{''}_{0}}(u)\label{eq:matrix}
\end{eqnarray}
which has the graphical representation given in Fig.(\ref{f4}).
It carries $N$ parameters $u_i$ from the bulk $S$-matrices and
additional ones from $R^a_{bc}$.
To form a commuting transfer
matrix out of ${\bf R}^a_{bc}$, like in the vertex case, one has to
combine it with another BYBE solution (denoted here as $\tilde{R}$) as follows
\begin{equation}
{\bf T}(u)_{a_0,\cdots,a_{N};a^{''}_0,\cdots,
a^{''}_N}\equiv \sum_{a,b,c}\tilde{R}^{a}_{cb}(u)
{\bf R}^a_{b,c}(u)_{a_0,\cdots,
a_N;a^{''}_0,\cdots,a^{''}_N}\;. \label{eq:transfer}
\end{equation}
Hence the transfer matrix ${\bf T}(u)$ is again an operator
acting on ${\cal P}_{N}$.

To show that
\[[{\bf T}(u),{\bf T}(u^{'})]=0\;,\]
one inserts four bulk $S$-matrices using the unitarity condition
\[\sum_{\alpha}S^{ac}_{b\alpha}(u^{'}+u) S^{\alpha c}_{bd}(-u^{'}-u)\propto
\delta_{ac}\;,\]
and the crossing-unitarity condition
\[\sum_{\alpha}S^{b\alpha}_{cd}(u^{'}+u)
S^{d\alpha}_{ab}(\gamma-u+\gamma-u^{'})\propto
\delta_{ac}\;,\]
into ${\bf T}(u){\bf T}(u^{'})$. Then, one uses the BYBE to permute the
${\bf R}$'s, and the $\tilde{R}$'s. Because of the argument
$\gamma-u+\gamma-u^{'}$ that appears in the crossing-unitarity condition,
one can take
\begin{equation}
\tilde{R}^{a}_{bc}(u)\equiv R^a_{bc}(\gamma-u)\label{eq:rbm}\;.
\end{equation}

\section{Diagonalization of the Transfer matrix}

So far, we managed to obtain solutions to the BYBE and construct the
corresponding commuting transfer matrix. It would be necessary
to diagonalize the transfer matrix in order to study the
statistical models given by these solutions. For applications
to field theory, diagonalization of the transfer matrix is also
needed in order to write down the thermodynamic Bethe ansatz equation.
For this purpose,
a systematic approach generalizing the algebraic Bethe ansatz for the case of
the periodic boundary condition has been devised in \cite{skly}.
However, the method
relies upon the conservation of the $S^z$ in the vertex language
and is thus applicable only to diagonal boundary scattering theories.
Therefore, we shall consider only the diagonal scattering solution
and adapt the algebraic Bethe ansatz method devised in \cite{skly},
along the line of \cite{devg}, to the SOS model with boundary.
The algebraic Bethe ansatz relies upon the existence of some pseudo-vacuum,
which in the vertex model, is a state with either all spins equal to $1/2$
or $-1/2$. In the SOS model, such a state corresponds to a
path which takes the form of a $45^{\rm o}$-oriented straight line in the
Bratteli diagram.
It is obvious that some heights on such a path,
for lattice size $N$ large enough ($>p$), have to exceed the
bounds $1$ or $p-1$. Therefore, this pseudo-vacuum
does not belong to the {\bf truncated} Bratteli diagram and
the algebraic Bethe ansatz that we are going to use is applicable to
the SOS model only.
For consistency, we have to assume that $\gamma/\pi$ is an irrational number.

\begin{figure}[htbp]
\centering
\input{face5}
\caption{\label{f5} Lattice generated by the transfer matrix }
\end{figure}
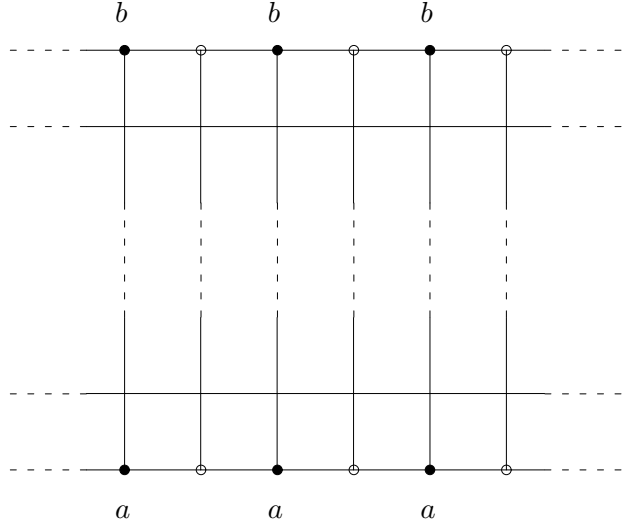

To diagonalize the transfer matrix given in eqn.(\ref{eq:transfer}), we
have to first write down the algebraic relations satisfied
by ${\bf R}^a_{bc}$.
As before, we express the operator ${\bf R}^a_{bc}$ as follows
\begin{eqnarray}
{\bf R}^{a}_{bc}(u)&=&\cR_N(u)\left(\frac{[b][c]}{[a][a]}\right)^{-u/2\gamma}
\prod_{j=2}^{N}\left(\frac{[a_j^{''}]}{[a_j]}\right)^{(u_j-u_{j-1})/2\gamma}
\left(\delta_{b\neq c}{\bf X}^{a}_{bc}(u)\right. \nonumber\\
&&\mbox{\hspace{2cm}}\left.+\delta_{bc}
(\delta_{b,a+1}{\bf U}_a(u)+\delta_{b,a-1}{\bf D}_a(u))\right)
\label{eq:operator}
\end{eqnarray}
where
\[\cR_N(u)=\cR(u)\prod_{i=1}^{N}{\cal U}(u-u_i){\cal U}(u+u_i)\;,\]
which, along with $q$-number factors, ensures the boundary crossing
and unitarity symmetry of ${\bf R}^a_{bc}$.
Here ${\bf X}^a_{bc},{\bf U}_a,{\bf D}_a$ are now
non-commutative operators that satisfy the algebraic
relations encoded by the BYBE. (see appendix)
Because $\tilde{R}^a_{bc}$ is diagonal, the commuting transfer matrix
can be written as
\begin{eqnarray}
{\bf T}(u)&=&\cR(\gamma-u)\cR_N(u)\prod_{j=1}^{N-1}
\left(\frac{[a_j^{''}]}{[a_j]}\right)^{(u_j-u_{j+1})/2\gamma}
\left(\frac{[b-1]}{[b]}U_{b-1}(\gamma-u){\bf U}_{b-1}(u)\right.\nonumber\\
&&\mbox{\hspace{2cm}}\left.+
\frac{[b+1]}{[b]}D_{b+1}(\gamma-u){\bf D}_{b+1}(u)\right)\;.\label{eq:comtran}
\end{eqnarray}
It is important to bear in mind that the RSOS heights
$a_{0},a_{0}^{''}$ on the ``bottom''
boundary (see Fig.(\ref{f4})) are set to be the same
since we consider only diagonal scattering.
Let us denote them as $a$, while the heights
on the other boundary are taken to be $b$ as evident from the above equation.
Hence the statistical model associated with the transfer matrix is
the one defined on a square lattice whose boundary heights are shown
in Fig.(\ref{f5}), where heights denoted by open-circles are summed over.

Consider the column of heights in Fig.(\ref{f4})
where $a_j=a+j$ and denote this state as $\omega_a^{a+N}$.
By construction, $\omega_a^{a+N}$ is an
eigenstate of ${\bf U}_{a+N-1}(u)$ and ${\bf D}_{a+N+1}(u)$
since the top and bottom heights of ${\bf U}_{a+N-1}(u)\omega_a^{a+N}$ and
${\bf D}_{a+N+1}(u)\omega_a^{a+N}$ are $a+N$ and $a$ respectively.
Furthermore, it is annihilated by ${\bf X}^{a+N+1}_{a+N+2,a+N}(u)$ due to the
constraint that neighboring heights differ by $\pm 1$.
Explicitly,
\begin{eqnarray}
{\bf U}_{a+N-1}(u)\omega_a^{a+N}&=&{\cal U}_{a}^{N}(u)\omega_a^{a+N}\nonumber\\
{\bf D}_{a+N+1}(u)\omega_a^{a+N}&=&{\cal D}_{a}^{N}(u)\omega_a^{a+N}\nonumber\\
{\bf X}^{a+N+1}_{a+N+2,a+N}(u)\omega_a^{a+N}&=&0
\end{eqnarray}
where the eigenvalues are given by
\begin{eqnarray}
{\cal U}_a^N(u)&=&\sum_{j=1}^{N}\left(\prod_{i=j+1}^{N}{\cal A}_{a+i}(u_i)
\right){\cal B}_{a+j}(u_j)\left(\prod_{k=1}^{j-1}{\cal C}_{a+k}(u_k)\right)
D_{a+1}(u)\nonumber\\
&&\mbox{}+\prod_{j=1}^{N}{\cal A}_{a+j}(u_j)U_{a-1}(u)\nonumber\\
{\cal D}_a^N(u)&=&\prod_{j=1}^{N}{\cal C}_{a+j}(u_i)D_{a+1}(u)
\end{eqnarray}
with
\[\begin{array}{lll}
{\cal A}_{a+i}(u_i)&\equiv&\frac{\textstyle \sin((a+i)\gamma)
\sin((a+i-2)\gamma)\sin(u-u_i)\sin(u+u_i)}
{\textstyle \sin^2((a+i-1)\gamma)}\\
{\cal B}_{a+i}(u_i)&\equiv&\frac{\textstyle\sin^2\gamma
\sin((a+i-1)\gamma+u-u_i)
\sin((a+i-1)\gamma+u+u_i)}{\textstyle\sin^2((a+i-1)\gamma)}\\
{\cal C}_{a+i}(u_i)&\equiv&\sin(\gamma-u+u_i)\sin(\gamma-u-u_i)
\end{array}\]
and $U_{a-1}(u),D_{a+1}(u)$ are given by
eqns.(\ref{eq:udsol1}),(\ref{eq:udsol2}).

The state ${\bf X}^{a+N-1}_{a+N-2,a+N}(\lambda)\omega_a^{a+N}$ is
non-zero, which
corresponds to a column of spins where the ``bottom'' and ``top'' spins
have heights $a$ and $a+N-2$ respectively.
So to obtain a state whose top spin has height
given by $b$, one has to act on $\omega_a^{a+N}$ successively
with $M\equiv (N+a-b)/2$ operators
${\bf X}^{a+N-i}_{a+N-i-1,a+N-i+1}(\lambda_i)\;\;,i=1,\cdots,M$.
We shall denote such a state as
\begin{equation}
\psi_a^b(\vec{\lambda})\equiv {\bf X}^{b+1}_{b,b+2}(\lambda_1)
\cdots{\bf X}^{a+N-1}_{a+N-2,a+N}(\lambda_M)\omega_a^{a+N}\label{eq:state}
\end{equation}
which is the Bethe ansatz state, and the set of parameters
$\vec{\lambda}\equiv (\lambda_1,\cdots,\lambda_{M})$ have to
satisfy some consistency
condition necessary for $\psi_a^b(\vec{\lambda})$ to be an eigenstate of the
transfer matrix. Notice that $M$ is always an integer fixed by the
heights $a,b$ and $N$.

The algebraic relations among ${\bf X}^a_{a\pm 1,a\mp 1}, {\bf U}_a,{\bf D}_a$
necessary for our purpose are (see Appendix)
\begin{eqnarray}
{\bf U}_{a-1}(u^{'}){\bf X}^{a+1}_{a,a+2}(u)&=&g_{1a}(u^{'},u)
{\bf X}^{a+1}_{a,a+2}(u){\bf U}_{a+1}(u^{'})
+g_{2a}(u^{'},u){\bf X}^{a+1}_{a,a+2}(u^{'}){\bf U}_{a+1}(u)\nonumber\\
&+&g_{3a}(u^{'},u){\bf X}^{a+1}_{a,a+2}(u^{'}){\bf D}_{a+3}(u)
+g_{4a}(u^{'},u){\bf X}^{a+1}_{a,a+2}(u){\bf D}_{a+3}(u^{'})\nonumber\\
{\bf D}_{a+1}(u^{'}){\bf X}^{a+1}_{a,a+2}(u)&=&f_{1a}(u^{'},u)
{\bf X}^{a+1}_{a,a+2}(u){\bf D}_{a+1}(u^{'})
+f_{2a}(u^{'},u){\bf X}^{a+1}_{a,a+2}(u^{'}){\bf D}_{a+1}(u)\nonumber\\
&+&f_{3a}(u^{'},u){\bf X}^{a+1}_{a,a+2}(u^{'}){\bf U}_{a+1}(u)\;,
\label{eq:nuse1}
\end{eqnarray}
where
\begin{eqnarray*}
g_{1a}(u^{'},u)&=&-\frac{\sin(a\gamma)\sin((a+1)\gamma)
\sin(\gamma-u^{'}+u)\sin(2\gamma-u^{'}-u)}
{\sin((a-1)\gamma)\sin((a+2)\gamma)\sin(\gamma-u^{'}-u)\sin(u^{'}-u)}\\
g_{2a}(u^{'},u)&=&\frac{\sin\gamma\sin((a+1)\gamma)
\sin(a\gamma+u^{'}-u)\sin(2\gamma-u^{'}-u)}
{\sin((a-1)\gamma)\sin((a+2)\gamma)\sin(\gamma-u^{'}-u)\sin(u^{'}-u)}\\
g_{3a}(u^{'},u)&=&-\frac{\sin\gamma\sin((a+1)\gamma)
\sin(a\gamma+u^{'}+u)\sin(2\gamma-u^{'}+u)}
{\sin((a-1)\gamma)\sin((a+2)\gamma)\sin(\gamma-u^{'}-u)\sin(u^{'}-u)}\\
g_{4a}(u^{'},u)&=&\frac{\sin\gamma\sin(2\gamma)
\sin(a\gamma+u^{'}+u)\sin((a+1)\gamma+u^{'}-u)}
{\sin((a-1)\gamma)\sin((a+2)\gamma)\sin(\gamma-u^{'}-u)\sin(u^{'}-u)}\\
f_{1a}(u^{'},u)&=&-\frac{\sin(u^{'}+u)
\sin(\gamma+u^{'}-u)}
{\sin(\gamma-u^{'}-u)\sin(u^{'}-u)}\\
f_{2a}(u^{'},u)&=&\frac{\sin\gamma\sin(u^{'}+u)
\sin((a+2)\gamma-u^{'}+u)}
{\sin((a+2)\gamma)\sin(\gamma-u^{'}-u)\sin(u^{'}-u)}\\
f_{3a}(u^{'},u)&=&-\frac{\sin\gamma\sin((a+2)\gamma-u^{'}-u)}
{\sin((a+2)\gamma)\sin(\gamma-u^{'}-u)}\\
\end{eqnarray*}

The action of the transfer matrix on the state
$\psi_a^b(\vec{\lambda})$ can be
evaluated by commuting ${\bf U}_{b-1}(u)$ and ${\bf D}_{b+1}(u)$
through ${\bf X}^{a+N-i}_{a+N-i-1,a+N+i-1}(\lambda_i)$. However, the presence
of the $g_{4a}(u^{'},u)$ term in the
first commutation relation complicates matters considerably.
Like the vertex case, it is desirable to define a new operator
\begin{equation}
\tilde{\bf U}_{a-1}(u)\equiv {\bf U}_{a-1}(u)+h_a(u){\bf D}_{a+1}(u)
\end{equation}
with
\[h_a(u)=-\frac{\sin\gamma\sin((a-1)\gamma+2u)}{\sin((a-1)\gamma)
\sin(\gamma-2u)}\;.\]
so that
the operators $\tilde{\bf U}_{a-1}(u),{\bf D}_{a+1}(u)$ satisfy
simplified relations
\begin{eqnarray}
\tilde{\bf U}_{a-1}(u^{'}){\bf X}^{a+1}_{a,a+2}(u)&=&\alpha_{1a}(u^{'},u)
{\bf X}^{a+1}_{a,a+2}(u)\tilde{\bf U}_{a+1}(u^{'})
+\alpha_{2a}(u^{'},u){\bf X}^{a+1}_{a,a+2}(u^{'})
\tilde{\bf U}_{a+1}(u)\nonumber\\
&&\mbox{}+\alpha_{3a}(u^{'},u){\bf X}^{a+1}_{a,a+2}(u^{'}){\bf D}_{a+3}(u)\\
{\bf D}_{a+1}(u^{'}){\bf X}^{a+1}_{a,a+2}(u)&=&\beta_{1a}(u^{'},u)
{\bf X}^{a+1}_{a,a+2}(u){\bf D}_{a+1}(u^{'})
+\beta_{2a}(u^{'},u){\bf X}^{a+1}_{a,a+2}(u^{'}){\bf D}_{a+1}(u)\nonumber\\
&&\mbox{}+\beta_{3a}(u^{'},u){\bf X}^{a+1}_{a,a+2}(u^{'})
\tilde{\bf U}_{a+1}(u)\;,
\end{eqnarray}
where
\begin{eqnarray*}
\alpha_{1a}(u^{'},u)&=&-\frac{\sin(a\gamma)\sin((a+1)\gamma)
\sin(\gamma-u^{'}+u)\sin(2\gamma-u^{'}-u)}
{\sin((a-1)\gamma)\sin((a+2)\gamma)\sin(\gamma-u^{'}-u)\sin(u^{'}-u)}\\
\alpha_{2a}(u^{'},u)&=&\frac{\sin\gamma\sin(a\gamma)
\sin((a+1)\gamma+u^{'}-u)\sin(2\gamma-2u)}
{\sin((a-1)\gamma)\sin((a+2)\gamma)\sin(\gamma-2u^{'})\sin(u^{'}-u)}\\
\alpha_{3a}(u^{'},u)&=&-\frac{\sin\gamma\sin(a\gamma)
\sin(2u)\sin(a\gamma+u^{'}+u)\sin(2\gamma-2u^{'})}
{\sin((a-1)\gamma)\sin((a+1)\gamma)\sin(\gamma-u^{'}-u)\sin(\gamma-2u^{'})}\\
\beta_{1a}(u^{'},u)&=&-\frac{\sin(u^{'}+u)
\sin(\gamma+u^{'}-u)}
{\sin(\gamma-u^{'}-u)\sin(u^{'}-u)}\\
\beta_{2a}(u^{'},u)&=&\frac{\sin\gamma\sin(2u)
\sin((a+1)\gamma-u^{'}+u)}
{\sin((a+1)\gamma)\sin(\gamma-2u)\sin(u^{'}-u)}\\
\beta_{3a}(u^{'},u)&=&-\frac{\sin\gamma\sin((a+2)\gamma-u^{'}-u)}
{\sin((a+2)\gamma)\sin(\gamma-u^{'}-u)}\;.\\
\end{eqnarray*}

Moreover the eigenvalue $\tilde{\cal U}_a^N(u)$ of $\tilde{\bf U}_{a-1}(u)$
on $\omega_a^{a+N}$ is much simplified and given by
\begin{equation}
\tilde{\cal U}_a^N(u)=-\frac{\sin(a\gamma)\sin(2u)
\sin(\xi_a+(a-1)\gamma+u)\sin(\xi_a+\gamma-u)}
{\sin((a-1)\gamma)\sin(\gamma-2u)\sin(\xi_a+a\gamma+u)
\sin(\xi_a-u)}\prod_{j=1}^N {\cal A}_{a+j}(u_j)U_{a-1}(u)
\end{equation}
where use has been made of the relation
\begin{equation}
{\cal B}_{a+i}(u_i)+h_{a+i}(u){\cal C}_{a+i}(u_i)=h_{a+i-1}(u)
{\cal A}_{a+i}(u_i).
\end{equation}

Another important relation we need is
\begin{equation}
{\bf X}^{a+1}_{a,a+2}(u){\bf X}^{a+3}_{a+2,a+4}(u^{'})=
{\bf X}^{a+1}_{a,a+2}(u^{'}){\bf X}^{a+3}_{a+2,a+4}(u)
\end{equation}
which implies that the Bethe ansatz state $\psi_a^b(\vec{\lambda})$ is a
symmetric function in the $\lambda_i$'s, which is useful in obtaining
the Bethe ansatz equation.

It is now straight forward to compute the action of
$\tilde{\bf U}_{b-1}(u),{\bf D}_{b+1}(u)$ on $\psi_a^b(\vec{\lambda})$ giving
\begin{equation}
\begin{array}{rl}
\tilde{\bf U}_{b-1}(u)\psi_a^b(\vec{\lambda})&=\alpha_{1b}(u,\lambda_1)
\alpha_{1b+2}(u,\lambda_2)\cdots\alpha_{1 a+N-2}(u,\lambda_{M})
\tilde{\cal U}_a^N(u)\psi_a^b(\vec{\lambda})\\
+&\left[\alpha_{2b}(u,\lambda_1)\alpha_{1b+2}(\lambda_1,\lambda_2)
\cdots\alpha_{1 a+N-2}(\lambda_1,\lambda_{M})
\tilde{\cal U}_a^N(\lambda_1)\right.\\
+&\left.\alpha_{3b}(u,\lambda_1)\beta_{1b+2}(\lambda_1,\lambda_2)
\cdots\beta_{1 a+N-2}(\lambda_1,\lambda_{M})
{\cal D}_a^N(\lambda_1)\right]\psi_a^b(\vec{\lambda};\lambda_1)\\
+&\cdots\\
+&\left[\alpha_{2b}(u,\lambda_M)\alpha_{1b+2}(\lambda_M,\lambda_2)
\cdots\alpha_{1 a+N-2}(\lambda_M,\lambda_{M-1})
\tilde{\cal U}_a^N(\lambda_M)\right.\\
+&\left.\alpha_{3b}(u,\lambda_M)\beta_{1b+2}(\lambda_M,\lambda_2)
\cdots\beta_{1 a+N-2}(\lambda_M,\lambda_{M-1})
{\cal D}_a^N(\lambda_M)\right]\psi_a^b(\vec{\lambda};\lambda_M)\\
\end{array}\end{equation}
and
\begin{equation}
\begin{array}{rl}
{\bf D}_{b+1}(u)\psi_a^b(\vec{\lambda})&=\beta_{1b}(u,\lambda_1)
\beta_{1b+2}(u,\lambda_2)\cdots\beta_{1 a+N-2}(u,\lambda_{M})
{\cal D}_a^N(u)\psi_a^b(\vec{\lambda})\\
+&\left[\beta_{2b}(u,\lambda_1)\beta_{1b+2}(\lambda_1,\lambda_2)
\cdots\beta_{1 a+N-2}(\lambda_1,\lambda_{M})
{\cal D}_a^N(\lambda_1)\right.\\
+&\left.\beta_{3b}(u,\lambda_1)\alpha_{1b+2}(\lambda_1,\lambda_2)
\cdots\alpha_{1 a+N-2}(\lambda_1,\lambda_{M})
\tilde{\cal U}_a^N(\lambda_1)\right]\psi_a^b(\vec{\lambda};\lambda_1)\\
+&\cdots\\
+&\left[\beta_{2b}(u,\lambda_M)\beta_{1b+2}(\lambda_M,\lambda_2)
\cdots\beta_{1 a+N-2}(\lambda_M,\lambda_{M-1})
{\cal D}_a^N(\lambda_M)\right.\\
+&\left.\beta_{3b}(u,\lambda_M)\alpha_{1b+2}(\lambda_M,\lambda_2)
\cdots\alpha_{1 a+N-2}(\lambda_M,\lambda_{M-1})
\tilde{\cal U}_a^N(\lambda_M)\right]\psi_a^b(\vec{\lambda};\lambda_M)\;,\\
\end{array}\end{equation}
where the state $\psi_a^b(\vec{\lambda};\lambda_i)$ is defined as in
eqn.(\ref{eq:state}) with $\lambda_i$ replaced by $u$. Combining the above
in the transfer matrix (see eqn.(\ref{eq:comtran})),
\begin{eqnarray}
{\bf T}(u)&\propto&\frac{\sin(2\gamma-2u)
\sin(\xi_b-u)\sin(\xi_b+b\gamma+u)D_{b+1}(\gamma-u)}
{\sin(\gamma-2u)\sin(\xi_b+\gamma-u)\sin(\xi_b+(b-1)\gamma+u)}
{\bf D}_{b+1}(u)\nonumber\\
&&\mbox{}+\frac{\sin((b-1)\gamma) U_{b-1}(\gamma-u)}{\sin(b\gamma)}
\tilde{\bf U}_{b-1}(u)\;,
\end{eqnarray}
the state $\psi_a^b(\vec{\lambda})$ is an
eigenstate of the transfer matrix if the coefficients of
$\psi_a^b(\vec{\lambda};\lambda_i),\;i=1,\cdots,M$ vanish, which
gives the following Bethe ansatz equation to be satisfied
for $\lambda_i$'s
\begin{equation}
\begin{array}{l} \displaystyle
\frac{\textstyle\sin(\xi_b-\gamma+\lambda_i)\sin(\xi_b+(b+1)\gamma-\lambda_i)
\sin(\xi_a+\gamma-\lambda_i)\sin(\xi_a+(a-1)\gamma+\lambda_i)}
{\textstyle\sin(\xi_b-\lambda_i)\sin(\xi_b+b\gamma+\lambda_i)
\sin(\xi_a+\lambda_i)\sin(\xi_a+a\gamma-\lambda_i)}\\ \\
= \displaystyle\prod_{\begin{array}{ccc} \scriptstyle
j&\scriptstyle =&\scriptstyle 1\\\scriptstyle j&\scriptstyle \neq&
\scriptstyle i\end{array}}^{M}
\frac{\textstyle\sin(\lambda_i+\lambda_j)\sin(\gamma
+\lambda_i-\lambda_j)}{\textstyle\sin(2\gamma-\lambda_i-\lambda_j)
\sin(\gamma-\lambda_i+\lambda_j)}\prod_{k=1}^{N}
\frac{\textstyle\sin(\gamma-\lambda_i+u_k)
\sin(\gamma-\lambda_i-u_k)}{\textstyle\sin(\lambda_i-u_k)
\sin(\lambda_i+u_k)}\;,\\ \hspace{8cm}\mbox{ for } i=1,\cdots,M
\end{array}\label{eq:bae}
\end{equation}
and the eigenvalue $\Lambda_a^b(u,\vec{u};\vec{\lambda})$ of the
transfer matrix
is given by
\begin{equation}\begin{array}{rcl}
\displaystyle\Lambda_a^b(u,\vec{u};\vec{\lambda})&=&\cR(\gamma-u)\cR_N(u)
U_{b-1}(\gamma-u)D_{a+1}(u)\\
&&\displaystyle\times\left(\frac{\sin(2\gamma-2u)\sin(\xi_b-u)
\sin(\xi_b+b\gamma+u)}{\sin(\gamma-2u)\sin(\xi_b-\gamma+u)
\sin(\xi_b+(b+1)\gamma-u)}\right.\\
&&\displaystyle\times\prod_{i=1}^{M}\frac{\sin(u+\lambda_i)
\sin(\gamma+u-\lambda_i)}{\sin(\lambda_i-u)
\sin(\gamma-u-\lambda_i)}\prod_{k=1}^N\sin(\gamma-u+u_k)\sin(\gamma-u-u_k)\\
&-&\displaystyle\frac{\sin(2u)\sin(\xi_a+\gamma-u)
\sin(\xi_a+(a-1)\gamma+u)}{\sin(\gamma-2u)\sin(\xi_a+u)
\sin(\xi_a+a\gamma-u)}\\
&&\displaystyle\left.\times\prod_{i=1}^{M}\frac{\sin(u+\lambda_i-2\gamma)
\sin(\gamma-u+\lambda_i)}{\sin(u-\lambda_i)\sin(\gamma-u-\lambda_i)}
\prod_{k=1}^N\sin(u-u_k)\sin(u+u_k)\right)\;\label{eq:eigen}
\end{array}\end{equation}

In the above analysis, $\xi_a$ and $\xi_b$ are respectively the free parameters
associated with the boundary $R$-matrices at the bottom and top boundaries.
They need not be related at all, hence more generally one should write them as
$\xi_{a}^{-}$ and $\xi_b^{+}$ to distinguish their origins.

\section{Discussion and open problems}
In this paper,  we present the general trigonometric RSOS/SOS solution to the
BYBE. By comparing them with the corresponding solution in the
vertex representation, one may be able to obtain useful information about
 the vertex-SOS transformation matrix for the BYBE \cite{zou}. Indeed,
for the diagonal solution, in the limit $\xi_a\rightarrow\pm i\infty$,
$U_{a-1}$ and $D_{a+1}$ are equal so they contribute as an overall factor
for the transfer matrix, since each bulk weight $S_{db}^{ac}$ is invariant
under the action of the quantum group ${\rm U}_q sl(2)$ symmetry
($q=e^{i\gamma}$), the transfer matrix possesses the quantum symmetry.
In this limit, the vertex-SOS transformation \cite{pas} is well known
and has a precise meaning in terms of the Clebsch-Gordan coefficients of
${\rm U}_q sl(2)$. Thus for generic $\xi_a$, the vertex-SOS transformation
can be considered as an extension of that of the ${\rm U}_q sl(2)$  case.

The SOS/RSOS solutions indicate that the diagonal solution is
not a special case of non-diagonal solution, in particular, there
is no way of adjusting the free parameters to make all the off-diagonal
scattering weights $X^a_{bc}$ vanish. The diagonal solution is the
most favorable case to be studied as we have demonstrated how
the transfer matrix of the SOS lattice built up from this solution
may be diagonalized. On the other hand, it is not obvious how such
method can be applied to the RSOS case. In the limit
$\pi/\gamma$ becomes a rational number, solutions to the Bethe ansatz
equation for the SOS model should contain those for the
RSOS. However, except at the special point where there
is a ${\rm U}_qsl(2)$ symmetry \cite{dev2}, it is not
clear how the RSOS solutions may be extracted. It would be an interesting
challenge to extend the idea to any $\xi_a$. The RSOS model
is indeed a very interesting case to consider; it has been shown that for
a different geometry where bulk faces are oriented at an angle $45^{\rm o}$
with respective to the boundaries, the
RSOS boundary condition where all spins have height $a$ corresponds in the
continuum limit to the boundary conformal state $|\tilde{h}_{1,a}\rangle$
and that with all boundary spins and their neighbors have
respectively heights $a,a+1$ corresponds to $|\tilde{h}_{a,1}\rangle$
\cite{bau}. The lattice model that we considered has a different gemeotry
from that in \cite{bau}, however, we believe that the difference is
not significant in the scaling limit. Then the former boundary condition
in fact can be obtained as the $\xi_a\rightarrow\pm i\infty$ limit
of the RSOS diagonal solution since the weights $U_{a-1},D_{a+1}$ become
the same. While the latter boundary condition can be obtained with
$\xi_a=u$. So for generic $\xi_a$, the RSOS diagonal solution is in
fact a mixture of the two above-mentioned boundary conditions and
it would be interesting to examine which boundary conformal state it
corresponds to in the scaling limit. Similarly, the non-diagonal solution,
which gives to some extent a free boundary-like condition, may correspond
to boundary conformal state $|\tilde{h}_{r,s}\rangle$ with $r,s\neq 1$
\cite{card} in the scaling limit.

As a scattering theory, our results should be related to
the conformal boundary conditions of the boundary minimal CFT
and perturbations by relevant operators
in the bulk and on the boundary. The non-diagonal solution, which has no
free parameter, can be interpreted as the bulk
$\Phi_{13}$ perturbation of the CFT with
free boundary condition where all possible spins are allowed on the boundary.
Additional integrable boundary perturbations introduce CDD-like factor
in the scattering amplitudes.
Each diagonal solution describes a perturbed CFT
with a fixed boundary condition where the boundary ${\bf B}_a$
has fixed spin $a$. The parameter $\xi_a$ in the solution should be related
to the coupling constant of the boundary perturbing field.
It is an open problem to relate our solutions to the specific
conformal boundary conditions and boundary perturbations of generic
minimal CFTs. The case $p=4$ has been analyzed in \cite{chim}.

One can also generalize our results to the coset CFTs
${\rm SU}(2)_k\otimes {\rm SU}(2)_l/{\rm SU}(2)_{k+l}$ perturbed by
the least relevant operator in the bulk and by some boundary fields.
The bulk-scattering matrices are given by
\[ S=S_{{\rm RSOS}(k+2)}\otimes S_{{\rm RSOS}(l+2)}\;,\]
where $S_{{\rm RSOS}(p)}$ is the RSOS $S$-matrix of the kinks.
In this theory, particles carry two sets of RSOS spins
and can be represented as
$\vert K_{ab}\rangle\otimes\vert K_{a^{'}b^{'}}\rangle$ \cite{bl}.
For the BYBE eqn.(\ref{eq:bybe}) with the above bulk scattering matrix,
the boundary $R$-matrix given by
\[ R=R_{{\rm RSOS}(k+2)}\otimes R_{{\rm RSOS}(l+2)}\;,\]
is a solution, where $R_{{\rm RSOS}(p)}$ is the $R$-matrix
given in eqn.(\ref{eq:boundary}). In particular, with $k=2$, this
is the $N=1$ super CFTs with boundary perturbed by the least relevant
operator and the $R$-matrix
is given by that of the tri-critical Ising model tensored with that of the
RSOS.

To answer some of the questions raised, finding the Bethe ansatz equation
and diagonalizing the transfer matrix for the RSOS case are the essential
first step. It is quite likely that one needs an alternative method
such as functional approach for this purpose.

\section*{Acknowledgment}
We thank A. LeClair who collaborated in the early stage of this work
and H. Saleur for many useful discussion, WMK also
acknowledges valuable discussion with P.Pearce. CA is supported
in part by KOSEF 95-0701-04-01-3, 961-0201-006-2
and BSRI 95-2427 and WMK by a grant from KOSEF
through CTP/SNU.

\section*{Note added in proof}
After completing this work, we learned that some of the results
presented in section 2, 3 have also been independently
obtained in \cite{pea,hou} as their trigonometric limit.

\section*{Appendix}
We present here the complete set of algebraic relations satisfied by
the operators ${\bf X}^{a}_{bc}(u)$, ${\bf U}_a(u)$, ${\bf D}_a(u)$.
These relations are obtained from the BYBE using
eqn.(\ref{eq:operator}) for the boundary scattering matrix.
Expanding the BYBE equation and considering the various allowed heights, we
get
\begin{equation}
\begin{array}{l}
c_{1a}(u_{+})c_{1a}(u_{-}){\bf U}_{a-1}{\bf U}_{a-1}^{'}
+c_{1a}(u_{-}){\bf X}^{a-1}_{a-2,a}{\bf X}^{'a-1}_{a,a-2}
+c_{3a}(u^{'},u){\bf U}_{a-1}{\bf D}_{a+1}^{'}\\
\mbox{}=c_{1a}(u_{+})c_{1a}(u_{-}){\bf U}_{a-1}^{'}{\bf U}_{a-1}
+c_{1a}(u_{-}){\bf X}^{'a-1}_{a-2,a}{\bf X}^{a-1}_{a,a-2}
+c_{3a}(u^{'},u){\bf D}_{a+1}^{'}{\bf U}_{a-1}
\end{array}
\end{equation}
\begin{equation}
\begin{array}{l}
c_{2a}(u_{+})c_{2a}(u_{-}){\bf D}_{a+1}{\bf D}_{a+1}^{'}
+c_{2a}(u_{-}){\bf X}^{a+1}_{a+2,a}{\bf X}^{'a+1}_{a,a+2}
+c_{3a}(u^{'},u){\bf D}_{a+1}{\bf U}_{a-1}^{'}\\
\mbox{}=c_{2a}(u_{+})c_{2a}(u_{-}){\bf D}_{a+1}^{'}{\bf D}_{a+1}
+c_{2a}(u_{-}){\bf X}^{'a+1}_{a,a+2}{\bf X}^{a+1}_{a+2,a}
+c_{3a}(u^{'},u){\bf U}_{a-1}^{'}{\bf D}_{a+1}
\end{array}
\end{equation}
\begin{equation}
\begin{array}{l}
c_{1a}f_{-}(u_{+}){\bf U}_{a-1}{\bf U}_{a-1}^{'}
+c_{2a}f_{+}(u_{-}){\bf U}_{a-1}{\bf D}_{a+1}^{'}
+f_{-}{\bf X}^{a-1}_{a,a-2}{\bf X}^{'a-1}_{a-2,a}\\
\mbox{}=c_{2a}f_{-}(u_{+}){\bf D}_{a+1}^{'}{\bf D}_{a+1}
+c_{1a}f_{+}(u_{-}){\bf U}_{a-1}^{'}{\bf D}_{a+1}
+f_{-}{\bf X}^{'a+1}_{a,a+2}{\bf X}^{a+1}_{a+2,a}
\end{array}
\end{equation}
\begin{equation}
\begin{array}{l}
c_{2a}f_{-}(u_{+}){\bf D}_{a+1}{\bf D}_{a+1}^{'}
+c_{1a}f_{+}(u_{-}){\bf D}_{a+1}{\bf U}_{a-1}^{'}
+f_{-}{\bf X}^{a+1}_{a,a+2}{\bf X}^{'a+1}_{a+2,a}\\
\mbox{}=c_{1a}f_{-}(u_{+}){\bf U}_{a-1}^{'}{\bf U}_{a-1}
+c_{2a}f_{+}(u_{-}){\bf D}_{a+1}^{'}{\bf U}_{a+1}
+f_{-}{\bf X}^{'a-1}_{a,a-2}{\bf X}^{a-1}_{a-2,a}
\end{array}
\end{equation}
\begin{equation}
\begin{array}{l}
c_{2a-2}(u_{+}){\bf X}^{a-1}_{a,a-2}{\bf X}^{'a-1}_{a-2,a}
+{\bf U}_{a-1}{\bf U}_{a-1}^{'}=c_{2a-2}(u_{+}){\bf X}^{'a-1}_{a,a-2}
{\bf X}^{a-1}_{a-2,a}\\
\mbox{     }+{\bf U}_{a-1}{\bf U}_{a-1}^{'}
\end{array}
\end{equation}
\begin{equation}
\begin{array}{l}
c_{1a+2}(u_{+}){\bf X}^{a+1}_{a,a+2}{\bf X}^{'a+1}_{a+2,a}
+{\bf D}_{a+1}{\bf D}_{a+1}^{'}=
c_{1a+2}(u_{+}){\bf X}^{'a+1}_{a,a+2}{\bf X}^{a+1}_{a+2,a}\\
\mbox{     }+{\bf D}_{a+1}^{'}{\bf D}_{a+1}
\end{array}
\end{equation}
\begin{equation}
{\bf X}^{a}_{a-1,a+1}{\bf X}^{'a+2}_{a+1,a+3}=
{\bf X}^{'a}_{a-1,a+1}{\bf X}^{a+2}_{a+1,a+3}
\end{equation}
\begin{equation}
\begin{array}{l}
{\bf X}^{a-1}_{a,a-2}{\bf D}_{a-1}^{'}
+c_{1a}(u_{+}){\bf U}_{a-1}{\bf X}^{'a-1}_{a,a-2}
=c_{1a}(u_{-}){\bf X}^{'a-1}_{a,a-2}{\bf D}_{a-1}\\
\mbox{}+c_{1a}(u_{+})c_{1a}(u_{-}){\bf U}_{a-1}^{'}
{\bf X}^{a-1}_{a,a-2}+c_{3a}(u^{'},u){\bf D}_{a+1}^{'}
{\bf X}^{a-1}_{a,a-2}
\end{array}
\end{equation}
\begin{equation}
\begin{array}{l}
{\bf X}^{a+1}_{a,a+2}{\bf U}^{'}_{a+1}
+c_{2a}(u_{+}){\bf D}_{a+1}{\bf X}^{'a+1}_{a,a+2}
=c_{2a}(u_{-}){\bf X}^{'a+1}_{a,a+2}{\bf U}_{a+1}\\
\mbox{ }+c_{2a}(u_{+})c_{2a}(u_{-}){\bf D}^{'}_{a+1}{\bf X}^{a+1}_{a,a+2}
+c_{3a}(u^{'},u){\bf U}_{a-1}^{'}
{\bf X}^{a+1}_{a,a+2}
\end{array}\label{eq:use1}
\end{equation}
\begin{equation}
\begin{array}{l}
c_{2a}(u_{-}){\bf U}_{a+1}{\bf X}^{'a+1}_{a+2,a}
+c_{2a}(u_{+})c_{2a}(u_{-}){\bf X}^{a+1}_{a+2,a}{\bf D}^{'}_{a+1}\\
\mbox{ }+c_{3a}(u^{'},u){\bf X}^{a+1}_{a+2,a}{\bf U}_{a-1}^{'}
={\bf U}^{'}_{a+1}{\bf X}^{a+1}_{a+2,a}+c_{2a}(u_{+}){\bf X}^{'a+1}_{a+2,a}
{\bf D}_{a+1}
\end{array}
\end{equation}
\begin{equation}
\begin{array}{l}
c_{1a}(u_{-}){\bf D}_{a-1}{\bf X}^{'a-1}_{a,a-2}
+c_{1a}(u_{+})c_{1a}(u_{-}){\bf X}^{a-1}_{a,a-2}{\bf U}^{'}_{a-1}\\
\mbox{ }+c_{3a}(u^{'},u){\bf X}^{a-1}_{a-2,a}{\bf D}_{a+1}^{'}
={\bf D}^{'}_{a-1}{\bf X}^{a-1}_{a-2,a}+c_{1a}(u_{+})
{\bf X}^{'a-1}_{a-2,a}{\bf U}_{a-1}
\end{array}
\end{equation}
\begin{equation}
\begin{array}{l}
f_{+}{\bf U}_{a-1}{\bf X}^{'a+1}_{a,a+2}=f_{-}{\bf X}^{'a+1}_{a,a+2}
{\bf U}_{a+1}+c_{1a}(u_{-})f_{+}{\bf U}_{a-1}^{'}{\bf X}^{a+1}_{a,a+2}\\
\mbox{ }+c_{2a}(u_{+})f_{-}{\bf D}^{'}_{a+1}{\bf X}^{a+1}_{a,a+2}
\end{array}
\end{equation}
\begin{equation}
\begin{array}{l}
f_{+}{\bf D}_{a+1}{\bf X}^{'a-1}_{a,a-2}=f_{-}{\bf X}^{'a-1}_{a,a-2}
{\bf D}_{a-1}+c_{2a}(u_{-})f_{+}{\bf D}_{a+1}^{'}
{\bf X}^{a-1}_{a,a-2}\\
\mbox{ }+c_{1a}(u_{+})f_{-}{\bf U}^{'}_{a-1}{\bf X}^{a-1}_{a,a-2}
\end{array}
\end{equation}
\begin{equation}
\begin{array}{l}
f_{-}{\bf U}_{a+1}{\bf X}^{'a+1}_{a+2,a}+c_{1a}(u_{-})f_{+}
{\bf X}^{a+1}_{a+2,a}{\bf U}_{a-1}^{'}+
c_{2a}(u_{+})f_{-}{\bf X}^{a+1}_{a+2,a}{\bf D}^{'}_{a+1}\\
\mbox{ }=f_{+}{\bf X}^{'a+1}_{a+2,a}{\bf U}_{a-1}
\end{array}
\end{equation}
\begin{equation}
\begin{array}{l}
f_{-}{\bf D}_{a-1}{\bf X}^{'a-1}_{a-2,a}+
c_{2a}(u_{-})f_{+}{\bf X}^{a-1}_{a-2,a}{\bf D}_{a+1}^{'}
+c_{1a}(u_{+})f_{-}{\bf X}^{a-1}_{a-2,a}{\bf U}^{'}_{a-1}\\
\mbox{ }=f_{+}{\bf X}^{'a-1}_{a-2,a}{\bf D}_{a+1}\label{eq:use2}
\end{array}
\end{equation}
where
\[\begin{array}{lll}
c_{1a}(u)&\equiv&\frac{\textstyle\sin\gamma\sin(a\gamma-u)}
{\textstyle\sin(a\gamma)\sin(\gamma-u)}\\
c_{2a}(u)&\equiv&\frac{\textstyle\sin\gamma\sin(a\gamma+u)}
{\textstyle\sin(a\gamma)\sin(\gamma-u)}\\
c_{3a}(u^{'},u)&\equiv&\frac{\textstyle\sin((a-1)\gamma)\sin((a+1)\gamma)
\sin(u^{'}+u)\sin(u^{'}-u)}
{\textstyle\sin^2(a\gamma)\sin(\gamma-u^{'}-u)\sin(\gamma-u^{'}+u)}\\
f_{\pm}&\equiv&\frac{\textstyle\sin(u^{'}\pm u)}
{\textstyle\sin(\gamma-u^{'}\mp u)}\\
u_{\pm}&\equiv&u^{'}\pm u\;.
\end{array}\]
Here again, we abbreviate
\[\begin{array}{ccc}
{\bf U}_a&\equiv& {\bf U}_a(u)\\
{\bf U}_a^{'}&\equiv& {\bf U}_a(u^{'})
\end{array}\] and similarly for ${\bf D}_a$, ${\bf X}^a_{bc}$.

Among them eqns.(\ref{eq:use1}),(\ref{eq:use2}) are of interest to
us, with the help of the latter the former can be turned into
the first equation in eqn.(\ref{eq:nuse1})
which is more convenient for the algebraic Bethe ansatz computation.

\end{document}

%% file: face1.tex
\setlength{\unitlength}{0.01in}
\begin{picture}(136,144)(0,-10)
\put(35.000,65.000){\arc{20.000}{4.7124}{7.8540}}
\path(35.000,55.000)(40.000,56.340)(43.660,60.000)
	(45.000,65.000)(43.660,70.000)(40.000,73.660)
	(35.000,75.000)
\put(25,65){\circle*{5}}
\put(65,105){\circle*{5}}
\put(105,65){\circle*{5}}
\put(65,25){\circle*{5}}
\path(65,105)(25,65)(65,25)
	(105,65)(65,105)
\put(0,65){\makebox(0,0)[lb]{$a$}}
\put(120,65){\makebox(0,0)[lb]{$c$}}
\put(65,120){\makebox(0,0)[lb]{$b$}}
\put(65,0){\makebox(0,0)[lb]{$d$}}
\put(50,60){\makebox(0,0)[lb]{\footnotesize $u-u^{'}$}}
\end{picture}

%% file: face2.tex
\setlength{\unitlength}{0.01in}
\begin{picture}(166,101)(0,-10)
\put(85,65){\circle*{5}}
\put(25,5){\circle*{5}}
\put(145,5){\circle*{5}}
\path(25,5)(85,65)(145,5)(25,5)
\spline(40,20)
(45,15)(45,5)
\put(85,80){\makebox(0,0)[lb]{$b$}}
\put(0,5){\makebox(0,0)[lb]{$a$}}
\put(55,10){\makebox(0,0)[lb]{$u$}}
\put(160,5){\makebox(0,0)[lb]{$c$}}
\end{picture}

%% file: face3.tex
\setlength{\unitlength}{0.008in}
\begin{picture}(638,181)(0,-10)
\put(25.000,25.000){\arc{22.361}{5.1760}{6.7468}}
\path(35.000,20.000)(36.160,25.670)(34.330,31.160)(30.000,35.000)
\put(125.000,25.000){\arc{22.361}{5.1760}{6.7468}}
\path(135.000,20.000)(136.160,25.670)(134.330,31.160)
	(130.000,35.000)
\put(64.167,95.833){\arc{21.730}{5.2791}{7.7772}}
\path(65.000,85.000)(70.088,86.723)(73.729,90.674)
	(75.032,95.886)(73.679,101.085)(70.000,105.000)
\put(177.500,52.500){\arc{35.355}{4.5705}{6.4251}}
\path(195.000,50.000)(194.931,55.442)(193.210,60.605)
	(190.000,65.000)(185.605,68.210)(180.442,69.931)
	(175.000,70.000)
\put(392.500,102.500){\arc{35.355}{6.1413}{7.9959}}
\path(390.000,85.000)(395.442,85.069)(400.605,86.790)
	(405.000,90.000)(408.210,94.395)(409.931,99.558)
	(410.000,105.000)
\put(445.000,57.500){\arc{25.000}{4.7124}{6.9267}}
\path(455.000,50.000)(457.249,55.009)(457.135,60.498)
	(454.680,65.409)(450.357,68.794)(445.000,70.000)
\put(390.000,25.000){\arc{22.361}{5.1760}{6.7468}}
\path(400.000,20.000)(401.160,25.670)(399.330,31.160)
	(395.000,35.000)
\put(525.000,25.000){\arc{22.361}{5.1760}{6.7468}}
\path(535.000,20.000)(536.160,25.670)(534.330,31.160)
	(530.000,35.000)
\put(20,20){\circle*{5}}
\put(100,20){\circle*{5}}
\put(60,100){\circle*{5}}
\put(140,140){\circle*{5}}
\put(260,20){\circle*{5}}
\put(180,60){\circle*{5}}
\put(220,100){\circle*{5}}
\put(360,20){\circle*{5}}
\put(400,100){\circle*{5}}
\put(440,60){\circle*{5}}
\put(480,140){\circle*{5}}
\put(520,20){\circle*{5}}
\put(560,100){\circle*{5}}
\put(600,20){\circle*{5}}
\path(60,100)(100,20)(180,60)
	(140,140)(60,100)
\path(60,100)(20,20)(100,20)(60,100)
\path(560,100)(520,20)(440,60)
	(480,140)(560,100)
\path(520,20)(360,20)(400,100)
	(480,140)(440,60)(360,20)
\path(560,100)(600,20)(520,20)(560,100)
\path(100,20)(260,20)(220,100)
	(140,140)(180,60)(260,20)
\put(35,105){\makebox(0,0)[lb]{$a$}}
\put(95,-5){\makebox(0,0)[lb]{$b^{'}$}}
\put(0,-5){\makebox(0,0)[lb]{$b$}}
\put(270,-5){\makebox(0,0)[lb]{$b^{''}$}}
\put(145,65){\makebox(0,0)[lb]{$a^{'}$}}
\put(130,160){\makebox(0,0)[lb]{$c$}}
\put(230,105){\makebox(0,0)[lb]{$a^{''}$}}
\put(85,90){\makebox(0,0)[lb]{\tiny$u^{'}+u$}}
\put(190,75){\makebox(0,0)[lb]{\tiny$u^{'}-u$}}
\put(475,160){\makebox(0,0)[lb]{$c$}}
\put(370,105){\makebox(0,0)[lb]{$a$}}
\put(570,105){\makebox(0,0)[lb]{$a^{''}$}}
\put(610,-5){\makebox(0,0)[lb]{$b^{''}$}}
\put(520,-5){\makebox(0,0)[lb]{$b^{'}$}}
\put(345,-5){\makebox(0,0)[lb]{$b$}}
\put(415,90){\makebox(0,0)[lb]{\tiny$u^{'}-u$}}
\put(470,60){\makebox(0,0)[lb]{\tiny$u^{'}+u$}}
\put(410,25){\makebox(0,0)[lb]{\tiny$u^{'}$}}
\put(150,25){\makebox(0,0)[lb]{\tiny$u^{'}$}}
\put(50,30){\makebox(0,0)[lb]{\tiny$u$}}
\put(550,30){\makebox(0,0)[lb]{\tiny$u$}}
\put(415,60){\makebox(0,0)[lb]{$a^{'}$}}
\put(300,80){\makebox(0,0)[lb]{\Large$=$}}
\end{picture}

%% file: face4.tex
\setlength{\unitlength}{0.0125in}
\begin{picture}(145,307)(0,-10)
\put(65,265){\circle*{5}}
\put(105,225){\circle*{5}}
\put(25,225){\circle*{5}}
\put(65,205){\circle*{5}}
\put(105,165){\circle*{5}}
\put(25,165){\circle*{5}}
\put(65,105){\circle*{5}}
\put(105,65){\circle*{5}}
\put(25,65){\circle*{5}}
\put(65,45){\circle*{5}}
\put(105,5){\circle*{5}}
\put(25,5){\circle*{5}}
\path(25,225)(65,265)(65,205)
	(25,165)(25,225)
\path(105,225)(65,265)(65,205)
	(105,165)(105,225)
\path(105,165)(105,145)
\path(65,205)(65,185)
\path(25,65)(65,105)(65,45)
	(25,5)(25,65)
\path(105,65)(65,105)(65,45)
	(105,5)(105,65)
\path(25,85)(25,65)
\path(105,85)(105,65)
\path(25,5)(105,5)
\dottedline{5}(25,145)(25,85)
\dottedline{5}(105,145)(105,85)
\path(65,125)(65,105)
\dottedline{5}(65,185)(65,125)
\path(25,165)(25,145)
\spline(35,15)
(40,10)(40,5)
\spline(65,215)
(75,210)(75,195)
\spline(65,115)
(75,110)(75,95)
\spline(65,55)
(75,50)(75,35)
\spline(35,235)
(35,220)(25,215)
\spline(35,175)
(35,160)(25,155)
\spline(35,75)
(35,60)(25,55)
\put(65,280){\makebox(0,0)[lb]{\tiny$a_N^{'}$}}
\put(45,105){\makebox(0,0)[lb]{\tiny$a_{1}^{'}$}}
\put(45,42){\makebox(0,0)[lb]{\tiny$a_{0}^{'}$}}
\put(5,230){\makebox(0,0)[lb]{\tiny$a_N$}}
\put(115,230){\makebox(0,0)[lb]{\tiny$a_N^{''}$}}
\put(115,160){\makebox(0,0)[lb]{\tiny$a_{N-1}^{''}$}}
\put(0,160){\makebox(0,0)[lb]{\tiny$a_{N-1}$}}
\put(40,205){\makebox(0,0)[lb]{\tiny$a_{N-1}^{'}$}}
\put(76,207){\makebox(0,0)[lb]{\tiny$u+u_N$}}
\put(38,220){\makebox(0,0)[lb]{\tiny$u-u_N$}}
\put(115,60){\makebox(0,0)[lb]{\tiny$a_1^{''}$}}
\put(115,0){\makebox(0,0)[lb]{\tiny$a_{0}^{''}$}}
\put(5,0){\makebox(0,0)[lb]{\tiny$a_{0}$}}
\put(5,60){\makebox(0,0)[lb]{\tiny$a_1$}}
\put(35,55){\makebox(0,0)[lb]{\tiny$u-u_1$}}
\put(78,45){\makebox(0,0)[lb]{\tiny$u+u_1$}}
\put(45,10){\makebox(0,0)[lb]{\tiny$u$}}
\put(76,105){\makebox(0,0)[lb]{\tiny$u+u_{2}$}}
\end{picture}

%% file: face5.tex
\setlength{\unitlength}{0.01in}
\begin{picture}(320,284)(0,-10)
\put(180,245){\circle{5}}
\put(220,245){\circle*{5}}
\put(260,245){\circle{5}}
\put(140,245){\circle*{5}}
\put(60,245){\circle*{5}}
\put(100,245){\circle{5}}
\put(100,25){\circle{5}}
\put(60,25){\circle*{5}}
\put(180,25){\circle{5}}
\put(220,25){\circle*{5}}
\put(260,25){\circle{5}}
\put(140,25){\circle*{5}}
\path(60,245)(40,245)
\path(60,205)(40,205)
\path(60,65)(40,65)
\path(60,25)(40,25)
\path(260,205)(260,245)(280,245)
\path(260,65)(260,25)(280,25)
\path(260,165)(260,205)(280,205)
\path(260,105)(260,65)(280,65)
\dashline{4.000}(260,165)(260,105)
\path(220,205)(220,245)(260,245)
\path(220,65)(220,25)(260,25)
\path(220,165)(220,205)(260,205)
\path(220,105)(220,65)(260,65)
\dashline{4.000}(220,165)(220,105)
\path(180,205)(180,245)(220,245)
\path(180,65)(180,25)(220,25)
\path(180,165)(180,205)(220,205)
\path(180,105)(180,65)(220,65)
\dashline{4.000}(180,165)(180,105)
\path(140,205)(140,245)(180,245)
\path(140,65)(140,25)(180,25)
\path(140,165)(140,205)(180,205)
\path(140,105)(140,65)(180,65)
\dashline{4.000}(140,165)(140,105)
\path(100,205)(100,245)(140,245)
\path(100,65)(100,25)(140,25)
\path(100,165)(100,205)(140,205)
\path(100,105)(100,65)(140,65)
\dashline{4.000}(100,165)(100,105)
\path(60,205)(60,245)(100,245)
\path(60,65)(60,25)(100,25)
\path(60,165)(60,205)(100,205)
\path(60,105)(60,65)(100,65)
\dashline{4.000}(60,165)(60,105)
\dashline{4.000}(320,25)(280,25)
\dashline{4.000}(320,65)(280,65)
\dashline{4.000}(320,205)(280,205)
\dashline{4.000}(320,245)(280,245)
\dashline{4.000}(40,65)(0,65)
\dashline{4.000}(40,25)(0,25)
\dashline{4.000}(40,205)(0,205)
\dashline{4.000}(40,245)(0,245)
\put(55,260){\makebox(0,0)[lb]{$b$}}
\put(135,260){\makebox(0,0)[lb]{$b$}}
\put(215,260){\makebox(0,0)[lb]{$b$}}
\put(55,0){\makebox(0,0)[lb]{$a$}}
\put(135,0){\makebox(0,0)[lb]{$a$}}
\put(215,0){\makebox(0,0)[lb]{$a$}}
\end{picture}

%% file: byb.bbl
\begin{thebibliography}{99}
\bibitem{cher}I.V.Cherednik, Teor. Mat. Fiz. {\bf 61} (1984) 35.
\bibitem{skly}E.K.Sklyanin,  J. Phys. {\bf A21} (1988) 2375.
\bibitem{oth}L.Mezincescu and R.I.Nepomechie, Int. J. Mod. Phys.
{\bf A6} (1991) 5231.
\bibitem{zam}S.Goshal and A.B.Zamolodchikov, Int. J. Mod. Phys.
{\bf A9} (1994) 3841.
\bibitem{ryu}P.P.Kulish and E.K.Sklyanin, J. Phys. {\bf A25} (1992) 5963;\\
P.P.Kulish, R.Sasaki and C.Schweibert, Lett. Math. Phys. {\bf 34} (1993) 286.
\bibitem{kul1}P.P.Kulish, {\em Quantum groups and quantum algebra as
symmetries of dynamical systems}, preprint YITP/K-959 (1991).
\bibitem{alek}A.Alekseev, L.Faddeev, M.Semenov-Tian-Shansky and
A.Volkov {\em The unraveling of the quantum group structure in WZNW theory},
 preprint CERN-TH-5981/91 (1991);\\
A.Alekseev, L.Faddeev and M.Semenov-Tian-Shansky, Commun. Math. Phys.
{\bf 149} (1992) 335.
\bibitem{naz}A.Molev, M.Nazarov and G.Olshanskii, {\em Yangian and classical
Lie algebras}, {\tt hep-th/9409025}.
\bibitem{kul}P.P.Kulish, {\em Yang-Baxter equation and reflection
equations in integrable models}, {\tt hep-th/9507070}.
\bibitem{pea}R.E.Behrend, P.A.Pearce and D.L.O'Brien, {\em
Interaction-Round-a-Face models with fixed boundary conditions:
The ABF fusion hierarchy}, {\tt hep-th/9507118}.
\bibitem{yung}C.M.Yung and M.T.Batchelor,  Nucl. Phys. {\bf B453} (1995) 552.
\bibitem{bl}C.Ahn,D.Bernard and A.LeClair, Nucl. Phys. {\bf B346} (1990) 409.
\bibitem{an}C.Ahn, Nucl. Phys. {\bf B354} (1991) 57.
\bibitem{zam1}Al.B.Zamolodchikov, Nucl. Phys. {\bf B358} (1991) 497.
\bibitem{fen}P.Fendley, Phys. Rev. Lett. {\bf 71} (1993) 2485.
\bibitem{abl}D.Bernard and A.LeClair, Nucl. Phys. {\bf B340} (1990)721.
\bibitem{meahn}C.Ahn and W.M.Koo, {\em Exact scattering matrices for
the supersymmetric sine-Gordon theory on a half line}, {\tt hep-th/9509056}.
\bibitem{resh}V.V.Bazhonov and N.Y.Reshetikhin, Int. J. Mod. Phys.
{\bf A4} (1989) 115.
\bibitem{devg}H.J.de Vega, Int. J. Mod. Phys. {\bf B5} (1990) 735.
\bibitem{zou}Y.K.Zhou, {\em Row transfer matrix functional relations for
Baxter's eight-vertex and six vertex models with open boundaries via
more general reflection matrices}, {\tt hep-th/9510095}.
\bibitem{gon}H.J.de Vega and A. Ruiz-Gonzalez, J. Phys. {\bf A26}
(1993) L519.
\bibitem{bax}R.J.Baxter, {\em Ann. of Phys.}, {\bf 76} (1973) 1.
\bibitem{men}W.M.Koo, unpublished note.
\bibitem{hou}H.Fan,B-Y.Hou and K-J Shi, {\em J. Phys.}, {\bf A28} (1995) 4743.
\bibitem{pas}V.Pasquier, Commun. Maths. Phys. {\bf 118} (1988) 355.
\bibitem{dev2}C.Destri and H.J.de Vega, Nucl. Phys. {\bf B385}
(1992) 361.
\bibitem{bau}H.Saleur and M.Bauer, Nucl. Phys. {\bf B320} (1989) 591;\\
Y.K. Zhou, Nucl. Phys. {\bf B453} (1995) 619.
\bibitem{card}J.L.Cardy, Nucl. Phys. {\bf B324} (1989) 581.
\bibitem{chim}L.Chim, {\em Boundary S-matrix for the tricritical
Ising model}, preprint RU-95-59, {\tt hep-th/9510008}.

\end{thebibliography}
